# A Survey on AI for 6G: Challenges and Opportunities

**CONSTANTINA CHATZIELEFTHERIOU (ORCID: 0009-0000-3058-9731), EIRINI LIOTOU**

[1]Department of Informatics and Telematics, Harokopio University, Athens, Greece

CORRESPONDING AUTHOR: Constantina Chatzieleftheriou (e-mail: constchatzi@hua.gr).

**ABSTRACT** As wireless communication evolves, each generation of networks brings new technologies that change how we connect and interact. Artificial Intelligence (AI) is becoming crucial in shaping the future of sixth-generation (6G) networks. By combining AI and Machine Learning (ML), 6G aims to offer high data rates, low latency, and extensive connectivity for applications including smart cities, autonomous systems, holographic telepresence, and the tactile internet. This paper provides a detailed overview of the role of AI in supporting 6G networks. It focuses on key technologies like deep learning, reinforcement learning, federated learning, and explainable AI. It also looks at how AI integrates with essential network functions and discusses challenges related to scalability, security, and energy efficiency, along with new solutions. Additionally, this work highlights perspectives that connect AI-driven analytics to 6G service domains like Ultra-Reliable Low-Latency Communication (URLLC), Enhanced Mobile Broadband (eMBB), Massive Machine-Type Communication (mMTC), and Integrated Sensing and Communication (ISAC). It addresses concerns about standardization, ethics, and sustainability. By summarizing recent research trends and identifying future directions, this survey offers a valuable reference for researchers and practitioners at the intersection of AI and next-generation wireless communication.

## I. Introduction

THE 6th generation (Sixth Generation (6G)) access network is expected to fulfill a new set of requirements and open unprecedented opportunities for immersive extended reality (Extended Reality (XR)), holographic communications, autonomous mobility, tactile internet, and the integration of terrestrial and non-terrestrial infrastructures [1]–[3]. Achieving these ambitious goals requires networks that are not only faster and more reliable, but also intelligent, adaptive, and sustainable [4]–[6].

Consequently, Artificial Intelligence (AI) is expected to become a native component of 6G systems, extending beyond its role in Fifth Generation (5G) networks as an auxiliary optimization tool. The convergence of AI and 6G enables a new network design paradigm that is self-manageable, self-performable, and more immersive, supporting real-time decision making, autonomous orchestration, and large-scale personalized services [7], [8].

Despite the significant potential of AI-driven 6G, research in this area remains fragmented and is expanding rapidly. Over the past five years, there has been a notable increase in studies exploring reinforcement learning (RL) for resource allocation, federated learning (FL) for privacy-preserving communications, explainable AI for reliable decision making, and, more recently, large language models (LLMs) for network automation [3], [9]–[18].

In parallel, efforts by industry groups such as the Open Radio Access Network (O-Radio Access Network (RAN)) Alliance, along with standardization bodies including 3GPP Release 18/19 and the ITU-T Focus Group on 6G, further emphasize the importance of integrating AI-native capabilities into future wireless networks [9], [19], [20].

However, the fast pace and broad scope of these changes present a challenge to both newcomers and experts. The research is spread across different fields, and there is no overall framework that clearly connects AI methods to the 6G system layers and services. Several survey articles have attempted to address parts of this problem. For example, Shi et al. [15] surveyed RL for 6G radio resource management, Porambage et al. [21] reviewed FL in wireless networks, and Chi et al. [22] investigated AI-driven RIS optimization. Broader overviews such as Kato et al. [23] and Yang et al. [2] provided early discussions on AI in 6G, but these either lack a taxonomy to avoid redundancy or do not reflect recent breakthroughs such as generative AI, and semantic communications. Consequently, the current body of survey literature is either too narrow in scope or outdated in coverage.

To our knowledge, no comprehensive and taxonomy-driven review has been published that bridges foundational AI paradigms with 6G system architecture while also synthesizing recent advances. This survey addresses this gap by offering a unified and structured review of AI for 6G. We propose a taxonomy that jointly classifies AI paradigms (RL, FL, explainable AI, large language models) and 6G system layers (physical, network/management, service, and cross-cutting domains). [12], [14], [15], [24]–[28]. We also review many publications published between 2018 and 2025, encompassing both foundational works and the latest state-of-the-art studies, thus providing readers with a balanced perspective. We introduce structured comparative tables that consolidate methodologies, datasets, performance metrics, and open challenges for each AI paradigm and application domain. Last but not least, we highlight emerging directions, including semantic communications, integrated sensing and communication (Integrated Sensing and Communication (ISAC)) and generative AI outlining their implications for 6G [29]–[31].

The remainder of this paper is organized as follows (Fig. 1). Section II reviews the state of the art and related surveys on AI for 6G. Section III highlights how this survey differs from prior work and outlines the adopted taxonomy. Section IV provides the necessary background on 6G architectures and the foundational AI concepts relevant to next-generation communications. Section V presents key AI techniques enabling 6G, including RL, FL, and deep learning. Section VI discusses the integration of AI and ML within 6G systems, emphasizing network optimization, management, and control. Section VII examines overarching challenges, open research issues, and potential solutions. Section VIII surveys practical applications and representative use cases of AI in 6G. Section IX concludes the paper. Finally, Section X outlines prospective avenues for future research in AI-driven 6G networks.

### A. Classification Methodology and Taxonomy

To provide a clear view and avoid repetition, this survey uses a taxonomy that organizes the role of AI in 6G along two dimensions: (i) the main AI concepts (ML, deep learning, RL, FL, and explainable AI) and (ii) the 6G system stack (physical, network/management, and service layers) [14], [20], [25], [32]–[34]. These are linked through common challenges such as scalability, security, sustainability, interoperability, and health considerations [21].

Fig. 2 illustrates this taxonomy Wachter et al. [35], which serves as the guiding framework for our survey. The figure illustrates the mapping between AI paradigms and the layered architecture of 6G systems, highlighting how different learning approaches enable intelligence across the network stack. Traditional machine learning techniques are primarily applied at the physical layer, supporting tasks such as channel estimation, reconfigurable intelligent surface (RIS) optimization, and terahertz (THz) communications. Deep learning and reinforcement learning play a central role at the network and management layer, where they enable data-driven orchestration, spectrum management, and control functions, including network analytics and decision-making frameworks. Federated learning is closely aligned with the service layer, facilitating privacy-preserving intelligence for applications such as the Internet of Things (IoT), extended reality (XR), and healthcare through distributed edge-based learning. Explainable AI acts as a cross-cutting paradigm, spanning multiple layers to enhance transparency, interpretability, and trust in AI-driven decisions. Across all layers, cross-cutting challenges—including security and privacy, scalability, energy efficiency, hardware constraints, interoperability, and health considerations—must be jointly addressed to enable the reliable and responsible deployment of AI-native 6G networks.To complement the taxonomy presented in Fig. 2, Table 1 summarizes how different AI paradigms map to 6G system layers and cross-cutting domains. This structured view avoids redundancy and guides the organization of the rest of this survey [36].

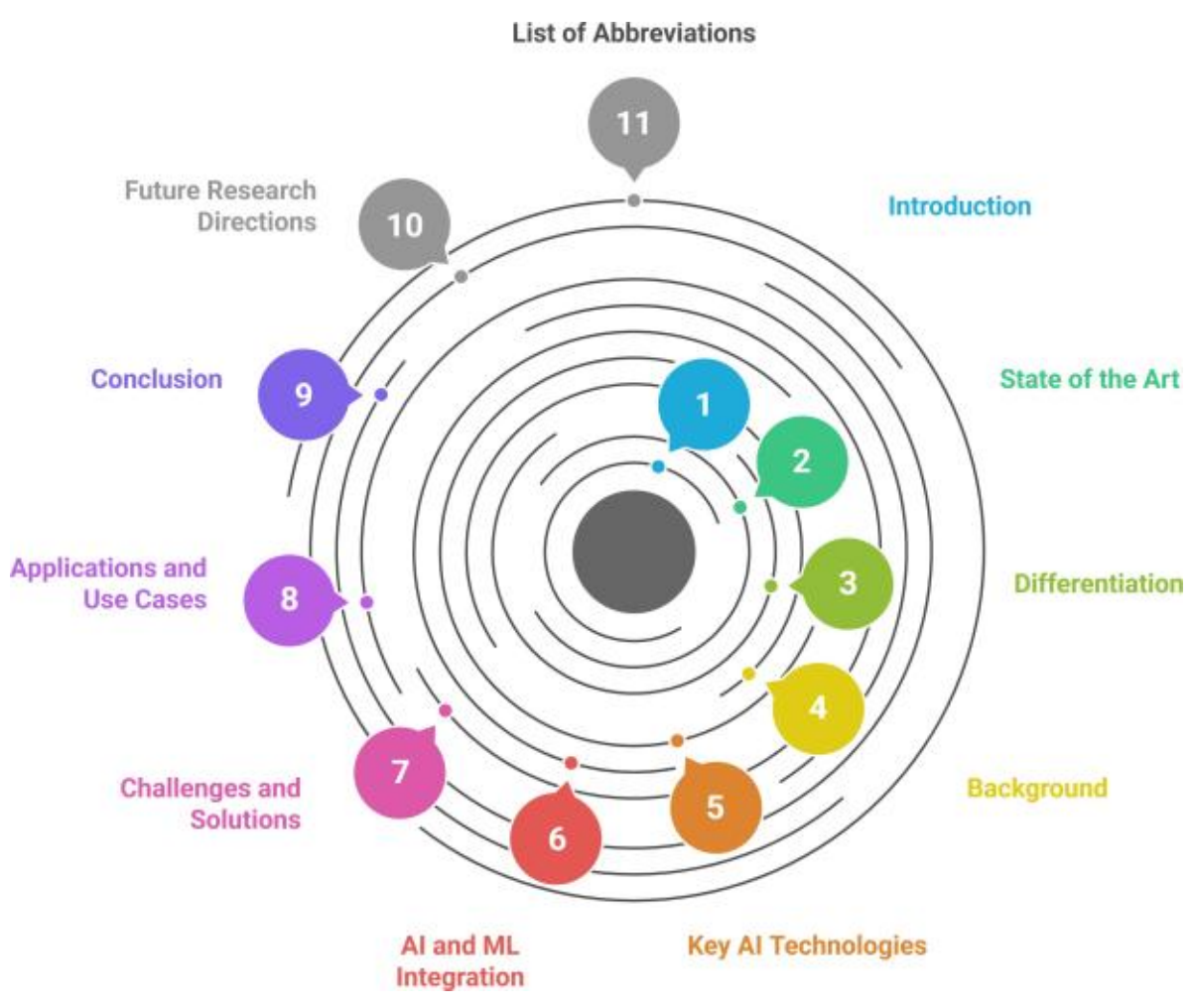

**FIGURE 1.** Survey Structure

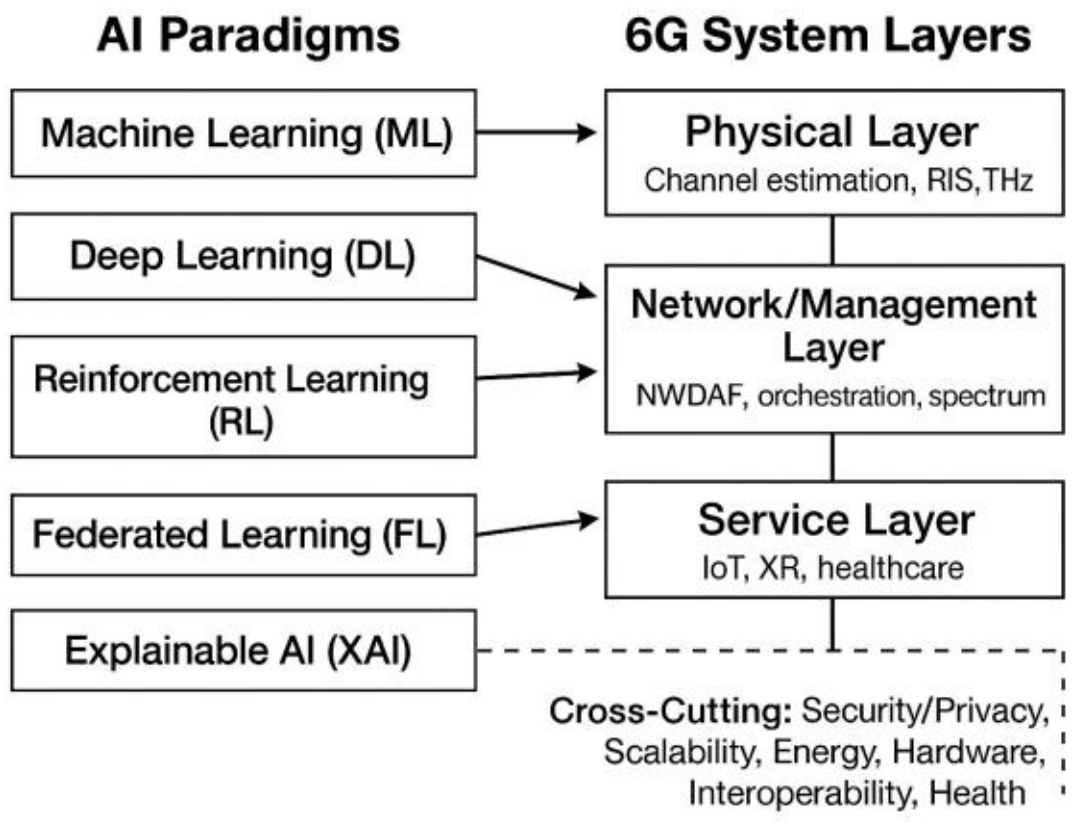

**FIGURE 2.** Taxonomy

## II. State of the Art on Related Surveys

Several recent surveys have explored the integration of artificial intelligence (AI) with next-generation wireless communication systems, primarily focusing on 5G and the transition toward 6G, as summarized in Table 2. For instance, Zhang *et al.* [40] provided an early overview of AI-driven techniques for 6G, highlighting representative use cases such as intelligent network control and spectrum management. Similarly, Jiang *et al.* [41] examined the role of edge intelligence and distributed learning in future communication networks, emphasizing challenges related to scalability and latency.

Other studies have focused on specific enabling technologies rather than the overall system perspective. These include AI-based resource allocation [42], AI-assisted physical layer design [43], and network automation enabled by Software-Defined Networking (SDN) and Network Function Virtualization (NFV) frameworks [44].

While these works have made valuable contributions to understanding how AI can enhance wireless communication systems, most of them either address isolated aspects of AI integration or remain confined to the 5G context. In contrast, the present survey provides a comprehensive examination of AI applications across the entire 6G ecosystem. It covers emerging research directions such as semantic communications, intelligent reflecting surfaces (IRS), joint communication and sensing, and AI-native network architecture design.

Moreover, this survey explicitly highlights cross-layer AI challenges, including data-driven optimization, algorithmic design, system-level orchestration, and security considerations. These aspects are critical for realizing fully intelligent and self-evolving 6G networks but are often treated separately in existing literature.

Overall, this work distinguishes itself from prior surveys by bridging the gap between enabling technologies and AI-native paradigms in 6G. By consolidating and extending insights from earlier studies, it provides a unified perspective on the challenges and opportunities that will shape future AI-empowered wireless communications.

From a different perspective, most existing surveys focus on only a single dimension of the AI–6G design space [53]. These include works centered on optimization techniques, deep learning models, O-RAN/RAN architectures, or individual network functions such as multiple access [13], [16], [17], [54]–[57]. As illustrated in Table 2, such studies lack a holistic cross-layer and service-oriented view.

In contrast, our survey adopts an AI-native, cross-layer perspective that explicitly connects 6G services and enablers, including URLLC, eMBB, mMTC, integrated sensing and communication (ISAC), and reconfigurable intelligent surfaces (RIS). We further incorporate the operational analytics stack, encompassing the Network Data Analytics Function (NWDAF) across the edge–core continuum, orchestration mechanisms, and key performance indicators (KPIs). Particular attention is given to representative AI paradigms such as deep learning (DL), reinforcement learning (RL) and multi-agent RL (MARL), federated learning (FL), and explainable AI (XAI).

In addition, this survey considers practical implementation constraints, societal implications, and standardization aspects. These include ethical considerations, Quality of Trust and Experience (QoTE), and health-related concerns associated with electromagnetic field (EMF) exposure and terahertz (THz) communications, which are often overlooked in technology-centric surveys.

### *Contributions of This Survey*

The main contributions of this survey are summarized as follows:

1) **AI-native architecture and NWDAF across the edge–core continuum:** We detail the capability, orchestration, and resource layers, and analyze how NWDAF evolves from its centralized role in 5G toward an edge–core continuum that supports real-time and predictive analytics, lifecycle scalability, and SDN/NFV integration (Fig. 1).
2) **Service-driven mapping between 6G use cases and AI methods:** We systematically map URLLC, eMBB, mMTC, and ISAC requirements to suitable AI techniques. Examples include edge AI and predictive maintenance for URLLC, federated learning and deep reinforcement learning (DRL) for mMTC interference management and spectrum sharing, and deep learning for holographic data compression and RIS control.
3) **Standards, governance, and QoTE/XAI considerations:** We examine interoperability challenges, standardized APIs, explainability requirements aligned with the General Data Protection Regulation (GDPR), and the QoTE perspective, which are often underrepresented in existing surveys.
4) **Hardware and edge orchestration requirements:** We analyze hardware–software co-design issues, AI accelerators, and edge server architectures, focusing on the stringent constraints imposed by AI-native 6G deployments.
5) **Spectrum sharing, coordination, and governance:** We review AI-driven approaches for spectrum sharing, dynamic routing, and cross-domain spectrum governance, leveraging techniques such as DRL, federated learning, and negotiation-based methods.

**TABLE 1. Taxonomy-guided coverage of AI in 6G**

| AI Paradigm | 6G Layer / Domain | Typical Tasks | Representative Works | Where |
|---|---|---|---|---|
| Deep Reinforcement Learning (DRL)/Multi-Agent RL (MARL) | Network / Management | Spectrum allocation, slicing, congestion control | Shi et al. [15], Yang et al. [2], Cao et al. [29] | Sec. V-A |
| Federated/Split Learning | Edge / Core | Privacy-preserving training, split inference | Porambage et al. [21], Mir et al. [1], Du et al. [3] | Sec. V-B |
| Trustworthy/Explainable Artificial Intelligence (XAI) | O-RAN / RAN Intelligent Controller (RIC) | Explainability, accountability, KPI attribution | Hossain et al. [4], Zhang et al. [5] | Sec. IV-A |
| RIS + ML | PHY / Environment | Beamforming, channel estimation, phase control | Chi et al. [22], Tshakwanda et al. [6] | Sec. V-C |
| Semantic Communications | PHY / Service | Task-oriented communications, meaning-aware encoding | Saruthirathanaworakun et al. [37], Maduranga et al. [38] | Sec. V-C |
| ISAC | Cross-domain | Joint sensing & communication, vehicular ISAC | Chataut et al. [7], Li et al. [11] | Sec. V-C |
| Large Language Models (LLMs)/Generative AI (GenAI) | Service / Management | Autonomic management, troubleshooting, planning | Noman et al. [12], Wang et al. [39] | Sec. IV-B |
| Digital Twins | Cross-layer | Network replicas, optimization, non-terrestrial networks (NTNs) | Tomkos et al. [8], Kato et al. [23] | Sec. VII-A |

**TABLE 2. Related Surveys on AI and 6G**

| Reference | Year | Venue | Focus | Notes |
|---|---|---|---|---|
| [45] | 2024 | IEEE Open Journal of the Communications Society | AI for 6G (General) | Comprehensive review of AI techniques for 6G; emphasizes taxonomy and open challenges. |
| [11] | 2024 | Intelligent and Converged Networks | AI for 6G (General) | Surveys ML for 6G wireless systems; lacks detailed experimental validation. |
| [19] | 2022 | IEEE Open Journal of the Communications Society | PHY / Waveform | Analyzes AI integration into next-gen communication; limited coverage of 6G-native aspects. |
| [46] | 2025 | IEEE Open Journal of the Communications Society | Network Slicing | Overviews AI-driven 6G architectures and enabling technologies; focuses mainly on conceptual design. |
| [47] | 2025 | Science China Information Sciences | AI for 6G (General) | Provides a broad overview of AI-assisted 6G frameworks; minimal discussion of implementation complexity. |
| [29] | 2024 | — | Joint Communication | Sensing Examines AI roles in 6G sensing and communication convergence; limited treatment of data-driven optimization |
| [48] | 2024 | IEEE Access | AI for 6G (General) | Surveys FL for 6G; highlights privacy and latency trade-offs without holistic integration. |
| [15] | 2023 | IEEE Communications Surveys &Tutorials | AI for 6G (General) | Reviews AI for semantic communication; primarily theoretical, with limited application-level insights. |
| [49] | 2023 | IEEE Communications Surveys & Tutorials | AI for 6G (General) | Focuses on ML for wireless networks; lacks emphasis on 6G-specific technologies. |
| [50] | 2021 | IEEE Communications Surveys & Tutorials | AI for 6G (General) | Provides taxonomy of ML use cases in 6G; missing discussion on network orchestration challenges. |
| [51] | 2021 | IEEE Communications Surveys & Tutorials | Joint Communication & Sensing | Covers joint communication and sensing in AI-enabled 6G; lacks cross-layer system analysis. |
| [52] | 2021 | IEEE Communications Surveys & Tutorials | Resource Allocation | Surveys enabling technologies for AI in 6G; conceptual overview without performance comparison. |

## III. Background

### A. 6G Overview

6G will bring over the current 5G in terms of performance and abilities. In this section, we present some key features and potential advances of 6G networks [55], [58]. The 6G network uses an AI-native logical architecture made up of three layers (Fig. 3). The first is the Capability Open Layer, which offers intelligent services with open capabilities that allow for seamless integration and collaboration. The second layer is the Function and Orchestration Management Layer. This layer manages collaborative AI control, resource orchestration, and policy enforcement [27], [59], [60]. Finally, there is the Resource Layer, which provides computational, spectrum, and data resources for distributed AI tasks [52], [58], [61].

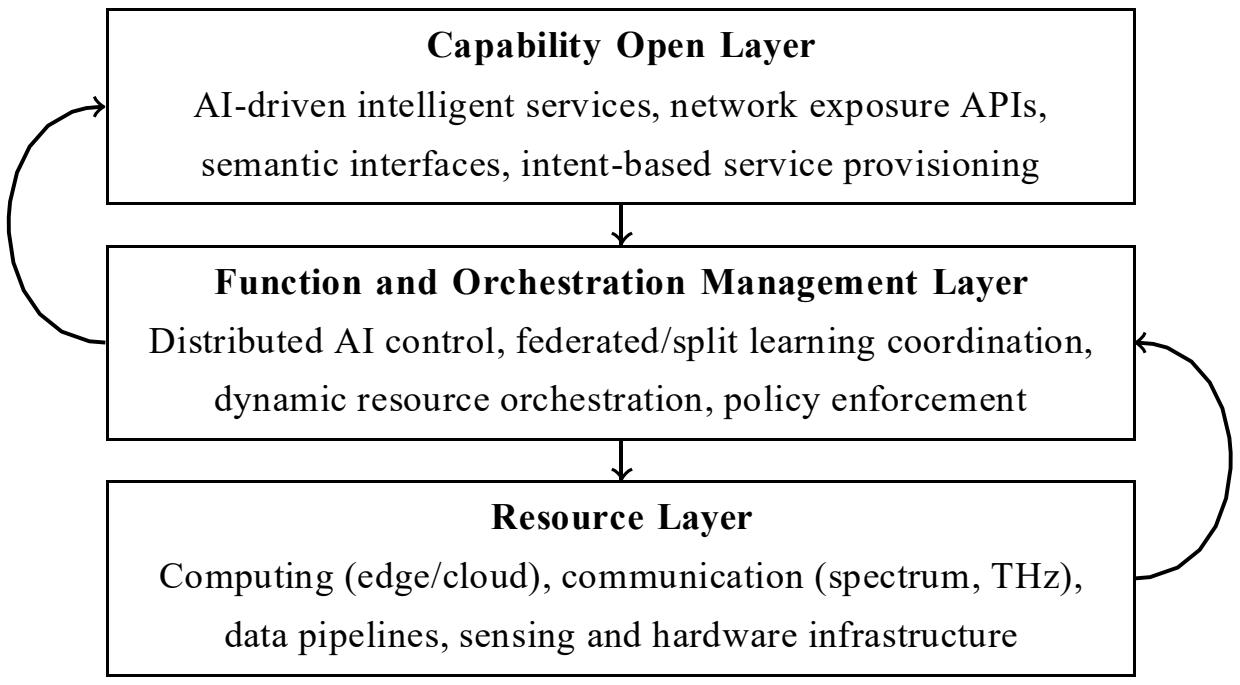

**FIGURE 3.** **Layered AI-native 6G architecture**

Further, integration of NWDAF enhances the capabilities to do real-time or predictive analytics related to operational optimizations Zhang et al. [53]. While in 5G this used to be a centralized function, in 6G, the NWDAF operates from edge to core and, therefore, facilitates localized decisions with global coordination [30]. NWDAF makes it possible to perform dynamic resource management, anomaly detection, and predictive maintenance using advanced AI techniques like deep learning and RL [18], [24], [56]. Some representative use cases are energy optimization, service quality assurance, and proactive adjustments by analyzing the behavior of users and traffic Hossain et al. [4]. However, scalability, compatibility with SDN and Network Virtualization Function (NFV), and lifecycle management of AI models are some of the key challenges for NWDAF in 6G [49], [62], [63]. High Data Transfer Rates in 6G networks are expected to achieve data transfer speeds that would probably exceed 10 terabits per second (Tbps), which would be an astonishing increase over the 10 Gbps of 5G networks [11], [64]. Such high-speed data transmission has great applications such as seamless 8K or 16K streaming, real-time holographic communication, and enhanced immersion in virtual and augmented realities [2], [3], [50]. Ultra-Low Latency is another important trait of the 6G networks which will provide a faster reduction of latencies to reach around 0.1 milliseconds (ms) [27], [65]. This is definitely better than the 1-ms latency requirement of 5G networks [66]. Such ultra-low latencies will be really crucial for mission-critical applications, such as remote surgery, autonomous vehicle control, and industrial automation, where real-time responsiveness is paramount [1], [4]. The extended coverage of 6G networks is so versatile that there might be deep-sea as well as space and underground environments within their coverage [67]. This also indicates the ability to develop applications like deep-sea exploration, space travel, or provide industrial internet at remote or difficult-to-reach locations. In addition, the extent of the improvement in coverage would connect and make the high-speed Internet more accessible in fairly neglected or rural areas, further contributing to the big picture of digital inclusion [2], [3]. Finally, with XR, users are expected to enjoy an enriched experiential environment in 6G networks [1], [4], [59], [68].

### *B. AI in Communications*

Given the extensive scalability of 6G networks in relation to users, devices, and services, the implementation of AI-driven automatic services is crucial [69]. The value created by enhanced feature extraction, decision making, and real-time optimizations is further supported by the integration of intelligent planes with data-centric designs [11]. Diagnostic analytics focus on identifying faults in the network, which allows the detection of anomalies, failures, and the root causes of network problems, which ultimately contributes to improved security and reliability [2], [3], [16], [55], [70], [71]. AI plays an important role in both 5G and 6G networks, with the focus of AI shifting from application-driven solutions in 5G to improving the overall design and management of the network in 6G, including reliable data sensing and efficient network management for applications such as connected autonomous vehicles [1], [4], [60]. AI will be essential in the development and implementation of 6G networks and their associated applications [61], [72], [73]. Examples include the use of generative adversarial networks (GANs) for predictive analysis and RL for real-time resource allocation [74]–[77]. There are various ways AI can contribute to 6G, with a common approach being the use of predictive, diagnostic, and descriptive analytics [3], [11]. AI will also play an important role in the control and orchestration of 6G networks [78], [79]. Furthermore, emerging frameworks such as Open RAN (O-RAN) will shape the future of Radio Access Networks (RAN) for 6G, with AI being integral to achieving key features within O-RAN, such as RIC frameworks [1], [80], [81]. Prescriptive analytics can help make informed decisions and predictions related to edge AI, including cache placement, AI model migration, dynamic scaling of network slices, adapting service function chains, and the optimal automatic allocation of resources, such as spectrum, cloud, and backhaul Hossain et al. [4]. AI-powered predictive analytics can analyze real-time data to forecast future events, including resource availability, user preferences, behavior, locations, and traffic patterns, allowing proactive adjustments to the network. These proactive measures can improve resource allocation, facilitate the deployment of security solutions, and support the premigration of edge services and AI models [2], [3], [82]. Descriptive analytics are based on historical data to improve the situational awareness of service providers and network operators. Applications include insights into user perspectives, channel conditions, traffic profiles, and overall network performance Hossain et al. [4]. In addition, managing, generating, and processing large volumes of data in real-time presents a complex challenge that requires scalable AI solutions [1], [2].

## IV. Key AI techniques for 6G

This section develops the technical foundations behind the AI paradigms used in 6G, organized by problem class and system-layer context Dang et al. [83]. For each paradigm, we: (i) formalize the problem and modeling choices, (ii) detail the algorithmic mechanisms that make the approach effective, (iii) discuss 6G-specific constraints that the literature addresses, (iv) compare representative studies, and (v) distill practical insights. Structured comparative tables are provided at the end of relevant subsections (Tables 3- 2).

### A. RL for Resource Allocation

Resource allocation in 6G spans spectrum sharing, power control, beam and user association, and slice admission, all under strict latency and reliability targets [70], [84], [85]. These optimization problems are typically nonconvex and highly environment-driven, which motivates the adoption of reinforcement learning (RL) approaches [86]–[88].

In a canonical formulation, a 6G controller, such as a near-real-time (RT) RIC xApp or a next-generation NodeB (gNB) scheduler, is modeled as an agent that observes the system state and selects control actions. The observed state may include channel quality indicators (CQI), buffer backlogs, and topology features, while actions correspond to physical resource block (PRB) assignment, power levels, or slice quotas. The agent receives a reward aligned with service-level objectives, including throughput, fairness, latency, and energy efficiency [79].

The algorithmic landscape reflects the diversity of RRM settings. Value-based methods, such as Q-learning and Deep Q-Networks (DQN), are effective when the action space is discrete and moderately sized. In contrast, policy-gradient and actor–critic methods, including Deep Deterministic Policy Gradient (DDPG), Twin Delayed Deep Deterministic Policy Gradient (TD3), and Proximal Policy Optimization (PPO), are better suited to continuous control problems such as power and rate adaptation [89].

Multi-agent reinforcement learning (MARL) extends these approaches to distributed decision-making across multiple cells or slices. Centralized training with decentralized execution (CTDE) is commonly adopted to handle interference coupling, while safety- and constraint-aware techniques, including Lagrangian RL, primal–dual updates, and reward shielding, are required to satisfy latency, power, and interference constraints.

Representative studies illustrate these trends. Shi [15] survey RL-based radio resource management (RRM) and report consistent gains over heuristic baselines. Yang et al. [2] demonstrate that MARL techniques such as Value Decomposition Networks (VDN) and QMIX improve spectral efficiency and fairness in distributed RRM, albeit at the cost of increased coordination overhead and credit-assignment complexity. Cao et al. [29] show that actor–critic MARL methods are competitive for dynamic spectrum and power allocation but may suffer from convergence and stability issues under non-IID traffic and high mobility unless enhanced with domain randomization or prioritized replay.

The literature further addresses several 6G-specific constraints. Near-real-time control deadlines on the order of 10–100 ms require low-inference-latency models, favoring shallow or quantized neural networks. Partial observability and channel state information (CSI) errors motivate partially observable Markov decision process (POMDP) formulations, recurrent policies, and adversarial training for robustness. Non-stationarity in traffic and interference calls for replay buffers, curriculum learning, and meta-RL techniques, while safety and service-level agreement (SLA) compliance necessitate constrained RL or offline-to-online learning pipelines [90], [91].

In practice, different RL approaches excel in different operational regimes. DQN and dueling DQN are efficient and stable for discrete action spaces with limited coupling, making them suitable for PRB scheduling. PPO, TD3, and DDPG perform well for continuous or smooth control tasks, with PPO often exhibiting greater stability under strict real-time constraints. MARL methods with CTDE, such as QMIX, scale to multi-cell scenarios but require careful management of credit assignment and inter-agent communication. Lagrangian actor–critic and Lyapunov-based methods are preferable when hard quality-of-service (QoS) or URLLC constraints must be strictly enforced.

Evaluation typically relies on simulation and prototyping toolchains such as ns-3/5G-LENA, MATLAB-based link- and system-level simulators, and open RAN stacks for hardware-in-the-loop testing. Performance assessment extends beyond throughput and fairness metrics, such as Jain's index, to include sample efficiency, SLA violations, and inference latency. Overall, RL and MARL consistently demonstrate gains for dynamic RRM by directly optimizing control policies based on observed system behavior. Actor–critic and MARL approaches align well with the decentralized nature of 6G networks, though they introduce challenges in coordination and stability. Constraint-aware design, robust state representation, and careful reward shaping emerge as critical factors for practical deployment. A comparative overview of methods, tools, and open challenges is provided in Table 3.

### B. Federated and Split Learning for Edge-Native Intelligence

#### 1) Motivation and Background

6G envisions ubiquitous intelligence at the edge, where data cannot always be centralized due to privacy, regulatory, or bandwidth constraints, as discussed by Wang et al. [39]. These limitations motivate learning paradigms that preserve data locality while enabling collaborative intelligence across distributed nodes [76], [92]. Consequently, distributed learning frameworks are widely considered core enablers of edge-native intelligence in 6G systems [49], [50].

**TABLE 3.** **Representative RL and MARL approaches for 6G resource allocation**

| Reference | Methodology | Scenario | Metrics | Key Insights / Limitations |
|---|---|---|---|---|
| [8], [23], [70], [84], [85] | RL-based optimization frameworks | General RRM (spectrum, power, slicing) | Throughput, latency, fairness | Formulates RRM as nonconvex, environment-driven problems; motivates RL adoption but lacks real-time constraints. |
| [37], [86]–[88] | Value-based RL (Q-learning, DQN) | Discrete RRM actions (PRB allocation) | Throughput, fairness | Effective for discrete action spaces; limited scalability for large or continuous control domains. |
| [1], [4], [89] | Policy-gradient / Actor–Critic (DDPG, TD3, PPO) | Power control, rate adaptation | Latency, energy efficiency | Better suited for continuous control; PPO shows improved stability under real-time constraints. |
| [2], [8], [23] | MARL (VDN, QMIX) with CTDE | Multi-cell and multi-slice RRM | Spectral efficiency, fairness | Improves coordination and fairness; introduces credit-assignment complexity and coordination overhead. |
| [1], [4], [29] | Actor–Critic MARL | Dynamic spectrum and power allocation | Convergence, stability | Competitive performance; sensitive to non-IID traffic and high mobility without replay or randomization. |
| [15], [23], [37] | Deep RL survey | RAN-level RRM | SE, latency, SLA violations | Reports consistent gains over heuristics; highlights deployment and sample-efficiency challenges. |
| [2], [4], [79] | Reward-aware RL | Energy-aware RRM | Energy efficiency, throughput | Explicit reward shaping enables energy–performance trade-offs; may increase policy complexity. |

### 2) Federated Learning and Split Learning Paradigms

To address these constraints, distributed learning paradigms such as Federated Learning (FL) and Split Learning (SL) have emerged as key enablers of edge-native intelligence.

FL enables collaborative training of a shared model across distributed clients, such as user equipments or edge nodes, by exchanging model updates instead of raw data. This approach inherently supports data sovereignty and privacy preservation.

In contrast, split learning partitions a neural network between devices and edge or cloud servers, thereby reducing memory requirements on constrained devices while limiting the exposure of sensitive representations, as shown by Huang et al. [65]. While both paradigms aim to preserve data locality, they exhibit different trade-offs in terms of communication, computation, and privacy [28].

### 3) Algorithmic Enhancements in Federated Learning

From an algorithmic perspective, a variety of mechanisms have been proposed to enhance FL performance under realistic 6G conditions. Federated Averaging (FedAvg) remains the baseline aggregation scheme; however, it is known to suffer from performance degradation under statistical heterogeneity.

To address non-IID data distributions, extensions such as Federated Proximal (FedProx) and adaptive client weighting have been proposed. Communication efficiency is improved through quantization, sparsification, and periodic aggregation, reducing uplink overhead without severely impacting convergence.

Robustness against poisoning and Byzantine attacks is enhanced using alternative aggregation rules, including coordinate-wise median and Krum, which mitigate the influence of malicious or unreliable clients, as analyzed by Mir et al. [1].

### 4) Algorithmic Design Choices in Split Learning

In SL, the placement of the cut layer governs the fundamental trade-off between device-side computation and communication overhead. Early cut layers reduce device computation but increase communication costs due to large activation tensors, while deeper splits shift computation to the device.

Pipeline parallelism and batch splitting techniques are commonly used to amortize training latency and improve throughput, as demonstrated by Du et al. [3]. Despite reduced raw data exposure, SL remains vulnerable to feature leakage across the split interface, motivating the use of additional privacy-preserving mechanisms [77], [86].

#### 5) Hybrid and Control-Oriented Learning Approaches

Beyond pure supervised learning, hybrid variants that integrate FL or SL with reinforcement learning (RL) are increasingly explored for control-oriented tasks in 6G systems, particularly for resource management and network optimization, as discussed by Huang et al. [65]. These approaches are especially relevant for dynamic resource allocation, scheduling, and mobility management, as shown by Lei et al. [93].

#### 6) Security and Privacy Challenges

Security and privacy considerations play a critical role in distributed learning for 6G. Porambage et al. [21] survey privacy-preserving FL architectures for beyond-5G and 6G systems, highlighting attack surfaces and system-level constraints.

Mir et al. [1] further demonstrate that both label-space and feature-space poisoning attacks can significantly degrade performance in the absence of robust aggregation or privacy mechanisms. Similarly, feature leakage in SL systems remains a concern unless additional defenses are deployed, as discussed by Li et al. [94].

#### 7) 6G-Specific System Constraints

Several 6G-specific constraints critically influence the applicability of FL and SL. Non-IID and imbalanced data across clients or network slices necessitate aggregation schemes that mitigate bias while ensuring fairness, as discussed by Liu et al. [72].

Client mobility and stragglers motivate asynchronous or partial aggregation strategies, as well as adaptive client selection policies, as shown by Abbasnasser et al. [74]. Limited communication budgets require compressed updates and reduced aggregation frequency, particularly when training must coexist with user traffic.

Privacy requirements are enforced through differential privacy and secure aggregation, providing quantifiable guarantees at the cost of reduced model accuracy, as analyzed by Wang et al. [82]. Governance and auditing considerations further motivate provenance tracking, logging, and rollback mechanisms to support trustworthy deployment.

#### 8) Paradigm Suitability Across Operational Contexts

Different learning paradigms are better suited to different operational contexts. FL with robust aggregation, quantized updates, and adaptive client sampling is effective under strict data sovereignty constraints and moderate compute availability.

SL or hybrid FL–SL approaches are advantageous for devices with very limited memory or when training deep models, particularly when split points are placed near downsampling layers to reduce activation size, as shown by Liu et al. [95]. Adversarial environments require combinations of robust aggregation, differential privacy, and secure aggregation, albeit with inherent accuracy–privacy trade-offs.

#### 9) Evaluation Metrics and Experimental Validation

Evaluation of FL and SL systems in 6G extends beyond accuracy to include convergence speed, communication volume per round, per-client energy consumption, and privacy loss. Experimental validation using edge prototypes and platforms such as srsRAN or OAI provides stronger evidence of practical feasibility, as demonstrated by Abbas et al. [75].

#### 10) Summary and Open Research Directions

Overall, FL and SL enable privacy-preserving and resource-aware edge learning, but their effectiveness hinges on addressing challenges such as non-IID data, stragglers, and communication constraints, as discussed by Cheng et al. [96]. A particularly promising direction lies in hybrid designs that combine FL or SL with RL for control tasks, alongside robust aggregation and communication-aware strategies [79], [80], [97].

A comparative overview of scenarios, methods, metrics, and limitations is provided in Table 4.

### *C. DL and GNNs for PHY/RIS*

#### 1) Role of Deep Learning in AI-Native PHY and RIS

Deep learning (DL) and graph neural networks (GNNs) are emerging as the cornerstones of the physical (PHY) and radio interface layers (RIS) in the AI-native 6G ecosystem, enabling context-aware, data-driven optimization across sensing, communication, and control functions.

Recent literature highlights the transformative role of deep learning in redefining how the PHY layer and reconfigurable intelligent surfaces (RIS) are modeled, optimized, and deployed. In the domain of semantic communications for extended reality (XR), several studies [26], [38], [39], [67], [99] demonstrate that DL-based pipelines trained on XR traffic traces can significantly enhance user quality of experience (QoE) and minimize end-to-end latency.

These frameworks leverage convolutional and attention-based neural architectures to extract semantically meaningful features from XR data streams, thereby enabling joint source–channel encoding that prioritizes task relevance over bit accuracy. Beyond compression and denoising, deep autoencoders and transformer variants have been utilized to perform task-oriented encoding, allowing communication systems to focus on reconstructing perceptual or decision-relevant information instead of raw signals. This paradigm shift from Shannon-driven communication to goal-driven semantic transmission yields substantial robustness gains under non-stationary environments and cross-modal interference.

**TABLE 4.** **Federated and split learning for 6G edge-native intelligence**

| Representative Work | Learning Paradigm | 6G Scenario | Evaluation Metrics | Key Insights / Limitations |
|---|---|---|---|---|
| Porambage et al. [21] | Federated Learning (survey) | Privacy-preserving wireless and edge networks | Accuracy, convergence, privacy budget | Provides a comprehensive survey of FL architectures and threat models; highlights vulnerability to poisoning attacks and communication overhead. |
| Mir et al. [1] | Federated Learning with robust aggregation and differential privacy | Secure FL for edge intelligence | Accuracy under attack, robustness | Demonstrates that FL performance degrades significantly under poisoning without robust aggregation; privacy–accuracy trade-offs remain critical. |
| Du et al. [3] | Split Learning | Resource-constrained edge devices | Memory footprint, latency | Shows that appropriate split-point selection reduces device memory and training latency, but intermediate feature leakage remains a concern. |
| Li et al. [87] | Federated Learning | Mobile edge computing for 6G | Communication cost, convergence speed | Highlights the impact of client mobility and non-IID data on FL convergence in MEC environments. |
| Vekariya et al. [55] | Hybrid FL–SL | Edge–cloud collaborative learning | Latency, communication overhead | Demonstrates that hybrid FL–SL schemes improve scalability, at the cost of increased system complexity. |
| Bi et al. [98] | Federated Learning | UAV-assisted 6G networks | Energy consumption, learning accuracy | Shows feasibility of FL in highly dynamic UAV scenarios, but notes sensitivity to stragglers and intermittent connectivity. |

### 2) DL- and GNN-Enabled Integrated Sensing and Communication

Integrated sensing and communication (ISAC) represents another key domain where DL and GNNs demonstrate pronounced advantages. Several studies [7], [31], [54], [73], [78] investigate the fusion of radar sensing and wireless communication through DL-based joint optimization frameworks.

In these approaches, neural networks are trained to perform beamforming, localization, and object detection simultaneously, learning unified feature representations that bridge sensing and communication modalities. In vehicular networks, such integration enables reliable environment perception under high mobility and rapidly varying channel conditions, while maintaining ultra-low latency and high reliability required by autonomous driving.

Performance is commonly evaluated using metrics such as F1 score, detection probability, and sensing latency, revealing that hybrid CNN–RNN models and attention-based fusion layers outperform traditional signal-processing-based ISAC pipelines, particularly under domain shift and noise corruption. Moreover, graph-based representations of vehicular topology allow GNNs to encode spatial dependencies among vehicles and base stations, improving coordination and prediction accuracy in distributed edge inference scenarios.

### 3) Edge-AI Optimization with DL and GNNs for RIS-Assisted Networks

At the network edge, DL- and GNN-based models have been applied to task-driven optimization in ISAC-enabled edge-AI frameworks. Representative works [6], [11], [45], [51], [59], [100] propose edge intelligence architectures in which learning components are co-located across edge nodes and RIS controllers to jointly manage task offloading, spectrum allocation, and interference mitigation.

These methods typically employ multi-task learning and knowledge distillation to adapt global models to heterogeneous hardware capabilities and channel conditions, balancing accuracy, energy efficiency, and training stability. Particular emphasis is placed on the interplay between inference latency and RIS configuration dynamics, as surface phase shifts must adapt in near-real time to channel variations.

To this end, lightweight GNNs and quantized neural architectures are developed to approximate optimal beamforming patterns with minimal computational overhead. Model compression, pruning, and federated training are further integrated to ensure scalability across large deployments involving thousands of distributed RIS tiles and edge nodes.

### 4) Theoretical Perspectives and Emerging Directions

From a theoretical standpoint, DL- and GNN-driven PHY/RIS solutions reflect a deeper convergence of model-based and data-driven design philosophies. While traditional PHY-layer optimization relies on analytically tractable propagation models, the 6G landscape increasingly embraces hybrid learning, where DL architectures embed inductive biases derived from physical principles such as channel symmetry, reciprocity, and spatial correlation.

GNNs, in particular, offer a natural abstraction for structured wireless environments, where nodes represent antennas, users, or RIS elements and edges capture dynamic channel interactions. Such representations enable scalable and interpretable learning over large-scale networks with limited pilot information.

At terahertz (THz) and sub-THz frequencies, where channel modeling becomes highly complex and nonlinear, deep neural networks have been applied to channel prediction, adaptive modulation, and RIS reflection design under severe path loss. As emphasized by Preskill et al. [101], quantum-inspired and physics-informed neural architectures may further bridge microscopic propagation models with macroscopic communication behavior, paving the way toward explainable and trustworthy PHY-layer intelligence in 6G.

#### 5) Summary and Implications for 6G PHY/RIS

In summary, deep learning- and GNN-based frameworks are poised to become essential enablers for 6G PHY and RIS design, unifying sensing, communication, and control under a single AI-native paradigm. The synergy between semantic understanding, graph-structured reasoning, and distributed edge computation enables adaptive, context-aware optimization across highly dynamic wireless environments. As PHY/RIS architectures evolve toward large-scale, programmable, and heterogeneous deployments, DL and GNNs provide the algorithmic foundation for autonomous, self-optimizing, and scalable 6G systems.

### D. Summary and Cross-Paradigm Insights

Across paradigms, three observations stand out. First, *constraints drive design*: near-RT deadlines, limited pilots, non-IID data, and safety requirements determine algorithm choice (PPO over DDPG for stability, robust aggregation for adversarial FL, GNNs for large arrays). Second, *hybridization is powerful*: RL+FL enables privacy-preserving control; model-based DL with GNN priors improves data efficiency and scalability; constrained RL integrates SLAs into learning. Third, *evaluation must be system-aware*: beyond task accuracy or reward, report inference latency, sample/round complexity, SLA violations, and energy. The tables in this section provide a consistent lens on methods, tooling, KPIs, and open challenges to support reproducible research and deployment in 6G Chi et al. [22].

## V. AI and ML Integration in 6G

#### 1) AI-Driven Beamforming and PHY Optimization

AI-driven methodologies have matured significantly beyond conventional beamforming based on classical signal processing. Machine learning (ML) techniques enable dynamic optimization of beam directions by adapting to user location, channel conditions, and interference levels, as discussed by Chataut et al. [7]. In this context, 6G systems increasingly rely on reinforcement learning (RL) combined with neural networks to maximize throughput and energy efficiency [6], [21].

Beamforming remains a core technology for enhancing directional transmission and reception, improving signal quality, reducing interference, and increasing network capacity. In 6G networks supporting ultra-reliable low-latency communication (URLLC), massive machine-type communication (mMTC), and enhanced mobile broadband (eMBB), adaptive beamforming is essential. Conventional beamforming techniques, however, exhibit limited adaptability to rapidly changing environments due to their reliance on static or semi-static signal models [102]. AI-based beamforming overcomes these limitations by leveraging data-driven optimization, enabling dynamic adjustment of beam patterns under heterogeneous and non-stationary conditions, as shown by Farhi et al. [103].

Table 5 compares key ML approaches for RIS and holographic MIMO in 6G, highlighting objectives, datasets, and evaluation metrics, while revealing persistent challenges in robustness under practical CSI errors.

#### 2) Limitations of Conventional Beamforming

Traditional beamforming approaches depend on fixed beam patterns derived from assumed network conditions, user distributions, and interference models. While effective in 4G and 5G systems, these methods face critical limitations in 6G scenarios [109]. Static beam patterns fail to adapt to mobility and dynamic traffic, real-time optimization incurs high computational complexity, and scalability becomes problematic as user density increases.

These limitations motivate the adoption of learning-based beamforming frameworks that continuously adapt to real-time measurements and historical data.

#### 3) Edge Intelligence and Federated Learning

Edge intelligence refers to the deployment of AI capabilities directly at network edge nodes, enabling local decision-making with minimal latency. This paradigm is critical for latency-sensitive applications such as autonomous driving and industrial automation. Federated learning (FL) plays a central role in edge intelligence by enabling collaborative model training without sharing raw data, as discussed by Porambage et al. [21].

FL supports privacy-preserving intelligence across heterogeneous devices while adapting to local network conditions [2]. Its integration into 6G enables scalable learning for applications such as smart cities and personalized healthcare [39], [45].

#### 4) AI-Enabled IoT Management

The Internet of Things (IoT) constitutes a foundational pillar of 6G, connecting billions of devices across smart cities, healthcare, industry, and agriculture. AI-driven IoT management enables predictive analytics for traffic forecasting and anomaly detection [8]. Deep learning models extract patterns from heterogeneous IoT data to optimize resource usage and ensure interoperability [38].

Predictive analytics forecast system behavior, while anomaly detection identifies deviations that may indicate faults, as demonstrated by Al et al. [84]. Deep learning architectures such as CNNs and RNNs further enhance pattern recognition and decision-making in IoT systems [88].

**TABLE 5.** RIS/Holographic MIMO with ML in 6G

| Paper | Objective | Method | Datasets / Sims | Metrics | Limitations |
|---|---|---|---|---|---|
| [5], [11], [12], [15], [22], [34], [46], [68], [104], [105] | RIS beamforming / channel estimation | DL-based (CNN, DNN, GNN) | DeepMIMO, ray-trace, synthetic | SE, EE | CSI errors, limited generalization |
| [6], [8], [16], [19], [20], [23], [32], [33], [48], [57], [75], [79], [106]–[108] | RIS for NTN / holographic MIMO | Hybrid DL + heuristics (DRL, meta, evolutionary) | NTN / holographic simulators | Coverage, EE, mobility, convergence | Channel dynamics, hardware limits, complexity |

#### 5) Federated Learning: Operation and Challenges

Federated learning is a distributed ML framework in which devices collaboratively train a shared model while retaining data locally. Each node performs local training, sends model updates to an aggregator, and receives a refined global model iteratively.

Despite its advantages, FL faces challenges in 6G environments. Communication overhead remains significant, mitigated through model compression and gradient quantization [89]. Device heterogeneity affects convergence, addressed via adaptive aggregation and personalized FL [90]. Privacy threats such as model inversion and poisoning persist, motivating defenses based on differential privacy and secure multi-party computation.

#### 6) AI for Spectrum Management and Network Control

AI-enabled spectrum monitoring improves utilization of scarce radio resources. Deep reinforcement learning (DRL) predicts spectrum availability and mitigates interference, enhancing throughput [37]. Speed-optimized LSTM models accelerate traffic forecasting in dynamic environments, supporting proactive resource allocation [6].

RL-based agents dynamically optimize load balancing, energy management, and beamforming under mobility and traffic fluctuations [4]. Bayesian filtering combined with RL further improves uncertainty handling in spectrum sensing and mobility prediction [29].

#### 7) Network Slicing, Hybrid Learning, and Model Partitioning

AI optimizes network slicing by dynamically allocating resources to meet diverse QoS requirements [5]. Federated edge learning balances exploration and exploitation to accelerate convergence in latency-critical applications [29].

Split learning (SL) distributes model computation between devices and servers to reduce resource burden, as discussed by Hossain et al. [4]. Hybrid FedSplit approaches combine FL and SL to achieve scalability and strong privacy guarantees.

Software-defined model partitioning dynamically allocates AI components across edge, fog, and cloud layers, enabling efficient utilization of computational resources [2]. Hybrid data and model parallelism further accelerates large-scale training.

#### 8) Ethics, QoTE, and Self-Supervised Learning

Quality of Trust and Experience (QoTE) emphasizes ethical and inclusive AI deployment in 6G. Fair resource allocation and bias mitigation are critical design goals [22]. Self-supervised learning enables model training without labeled data, improving robustness in data-scarce environments [21]. Generative adversarial networks (GANs) further support data synthesis and imbalance mitigation, enhancing performance in applications such as intrusion detection and medical imaging [26].

#### 9) AI and 6G Standardization Efforts

AI integration in 6G is increasingly reflected in standardization activities. The ITU-R has identified AI-native networking as a key capability for IMT-2030, emphasizing learning-driven PHY, MAC, and network management functions [110]. Similarly, 3GPP [111] has initiated studies on AI/ML for radio access networks, including channel state prediction, beam management, and energy optimization, as outlined in 3GPP TR 38.838 [112]

ETSI is also advancing standardization through its Experiential Networked Intelligence (ENI) framework, which defines cognitive closed-loop control architectures for AI-driven networks [113]. These efforts aim to ensure interoperability, explainability, and trustworthiness of AI components in future 6G deployments.

## VI. Challenges and Potential Solutions

### *A. Security and Privacy Challenges*

The infusion of artificial intelligence (AI) into 6G networks presents significant opportunities, but also introduces major challenges related to security and privacy. Several key innovations have already demonstrated their usefulness in essential aspects of 6G networking, offering exceptional reliability, lower latency, and secure and efficient transmission solutionsGao et al. [114]. However, these advancements may also introduce new privacy and security risks, particularly when AI systems operate over large-scale, sensitive datasets Vekariya et al. [55].

**TABLE 6. Security and Privacy Challenges and AI-Based Solutions in 6G Networks**

| Topic | Problem | AI-Based Solution |
|---|---|---|
| Privacy-Preserving Models in 6G | AI models rely on large-scale sensitive data, raising privacy risks and increasing computational overhead under real-time constraints. | Privacy-preserving AI using differential privacy and homomorphic encryption to protect confidential data. |
| Need for Privacy at 6G Scale | Massive data generation and processing amplify privacy risks while ultra-low latency requirements limit heavy protection mechanisms. | Optimized privacy-preserving AI techniques balancing privacy, accuracy, and real-time performance. |
| Differential Privacy | Noise injection degrades model accuracy, especially in high-precision and URLLC applications. | Differential privacy mechanisms integrated with FL to protect individual data during distributed training. |
| Homomorphic Encryption | Operations on encrypted data incur high computational cost and latency. | Homomorphic encryption enabling secure computation on encrypted data at edge and cloud environments. |
| Privacy-Preserving Overheads and Trade-offs | Privacy techniques increase computation and latency, conflicting with real-time 6G requirements. | AI model optimization to balance privacy guarantees, accuracy, and latency constraints. |
| AI-Driven Threats and Attack Surface | AI-enabled attacks (e.g., GAN-based phishing and malware) adapt to defenses and evade detection. | AI-based security mechanisms for detecting and mitigating intelligent cyberattacks. |
| Representative AI Attack Vectors | Highly realistic phishing, malware, and fake traffic bypass traditional security systems. | AI-driven detection and analysis of adaptive attack patterns. |
| Limits of Conventional Security | Static rules and signature-based systems fail against adaptive AI-driven attacks. | AI-enabled security approaches replacing static detection with adaptive learning-based defense. |
| AI-Enabled Countermeasures | Need for real-time detection and mitigation of intelligent and evolving attacks. | AI-driven intrusion detection, anomaly detection, and adversarial ML techniques. |
| Decentralized AI and DLT Security | Distributed AI systems face secure communication, data leakage, and malicious node risks. | Blockchain, DLTs, SMPC, FL, and differential privacy for secure decentralized AI operation. |
| FL Threats and Protections | Federated learning is vulnerable to model poisoning and inference attacks. | Secure aggregation and safe communication protocols for FL-based training. |
| Adversarial ML | Adversarial inputs can deceive AI models and disrupt 6G services. | Adversarial training, defensive distillation, and robust model design. |
| QoTE-Based Security Evaluation | Lack of metrics to assess trustworthiness of AI-based 6G systems. | QoTE-based evaluation considering reliability, transparency, and resilience. |
| AI for Detection, Authentication, and Access Control | Dynamic attacks challenge authentication and access control mechanisms. | AI-based anomaly detection, RL-based IDS, and continuous authentication. |
| Blockchain for Secure Data and Transactions | Need for transparent and secure data sharing with scalability and latency constraints. | Blockchain-based secure data management and transaction verification. |
| End-to-End Trust and Adversarial Readiness | Trust establishment across all 6G layers is vulnerable to manipulation and collusion. | AI-driven trust management and adversarially robust learning systems. |
| Endogenous Security and Privacy at the Edge | Distributed edge processing increases exposure to attacks and privacy leakage. | AI-enabled endogenous security with FL and secure multiparty computation at the edge. |

***Privacy-Preserving Models in 6G***

Most AI algorithms rely on large volumes of data, often including confidential user information. Privacy-preserving models aim to mitigate these concerns through techniques such as differential privacy and homomorphic encryption. Differential privacy introduces controlled noise into data values or query results, ensuring that individual data contributions cannot be inferred, while homomorphic encryption enables computations to be performed directly on encrypted data without requiring decryption.

Although these techniques enhance security, they also increase computational overhead. A representative example is the deployment of privacy-preserving AI models in medical systems, where sensitive patient data must be protected against unauthorized access. Key challenges in such systems include maintaining model accuracy and computational efficiency, particularly under the real-time constraints of 6G applications Chataut et al. [7].

***Need for Privacy at 6G Scale***

In 6G networks, the generation, transmission, and processing of massive volumes of sensitive user data significantly amplify privacy risks. As AI systems increasingly rely on large-scale datasets containing confidential information, privacy protection becomes a critical requirement. Privacy-preserving techniques have therefore emerged as central components in AI-enabled 6G systems.

However, these techniques often incur substantial computational overhead, which is especially problematic in 6G environments that demand ultra-low latency and high-speed processing. As a result, trade-offs between privacy, model accuracy, and real-time performance become more pronounced under 6G operational conditions.

***Differential Privacy***

Differential privacy provides a rigorous mathematical framework for ensuring strong privacy guarantees by limiting the impact of any single data point on the output of an analysis. This is typically achieved by injecting random noise into the data or query results, thereby making it statistically improbable to infer sensitive information about individuals.

Differential privacy is particularly effective when training AI models on datasets containing personal or sensitive data, such as in healthcare, finance, or user behavior analytics. In 6G networks, differential privacy can be applied to protect data generated by edge devices, IoT sensors, and mobile applications. Within federated learning (FL) settings, differential privacy prevents the disclosure of individual user data even when model updates are exchanged among distributed devices or centralized aggregators Shokri et al. [56].

***Homomorphic Encryption***

Homomorphic encryption allows computations to be performed directly on encrypted data, ensuring that sensitive information remains encrypted throughout the processing pipeline. This capability is particularly valuable when AI computations are carried out in untrusted or third-party environments, such as cloud or edge computing infrastructures.

In 6G networks, homomorphic encryption can secure data processing at the edge, where sensitive information is frequently exchanged between devices and edge servers. For example, in smart healthcare systems, patient data collected by wearable devices can be encrypted and processed by AI algorithms at the edge without decryption, ensuring privacy preservation and compliance with regulations such as the General Data Protection Regulation (GDPR) Brakerski et al. [108].

***Privacy-Preserving Overheads and Trade-offs***

Despite their advantages, privacy-preserving techniques introduce several challenges that must be carefully addressed in the 6G context. First, differential privacy increases computational complexity and can degrade model accuracy due to the injection of noise. Second, homomorphic encryption is computationally intensive, as operations on encrypted data require significantly more processing resources.

These limitations are particularly critical for 6G applications where ultra-low latency and real-time performance are non-negotiable. Noise introduced by differential privacy can adversely affect AI model accuracy, especially in high-precision applications such as medical diagnosis and autonomous driving. Consequently, identifying an optimal balance between privacy and accuracy remains a central challenge in the design of privacy-preserving AI systems Abadi et al. [77].

Furthermore, 6G networks are designed to support ultra-reliable low-latency communication (URLLC), requiring AI systems to operate in real time. Privacy-preserving techniques must therefore be optimized to meet stringent latency and reliability constraints without compromising either privacy guarantees or model performance Mao et al. [99].

***AI-Driven Threats and Attack Surface***

AI may also be misused to infiltrate and launch intelligent cyberattacks, such as phishing, malware creation, and evasive actions. A generative adversarial network would be able to create valid false data, capable of bypassing intrusion detection systems. In addition, AI-enabled tools could study and adapt themselves to the defense mechanisms within a particular network, which would make counteraction even more difficult. These threats can only be solved using AI-based techniques, which can identify and mitigate AI-driven attacks in 6G environments [7], [39], [45]. AI is to be used more and more in integrating through 6G networks up to its benefits and expose new kinds of risk as well. Such advanced techniques of AI include Generative Adversarial Networks (GANs) for creating genuine-false data, bypassing Intrusion Detection Systems (Intrusion Detection System (IDS)), and adapting them to defense mechanisms. This makes it very difficult to detect and mitigate such dynamic and intelligent attacks. Counter-measures against AI-based threats are necessary as they will detect and provide real-time mitigation against AI-initiated attacks to ensure the security and resilience of 6G networks. Such potential AI-powered cyber-attacks would pose a severe risk to all 6G networks that rely on ultra-reliable low-latency communication (URLLC), massive machine-type communications (mMTC), and enhanced mobile broadband (eMBB). The following are some important attack vectors of AI Yakubov et al. [115].

***Representative AI Attack Vectors***

AI can create highly personal and very believable phishing emails or messages by gleaning information from user behavior and preferences. Due to the vast scale of 6G networks, such attacks could be scaled up to affect multiple users simultaneously. It also allows the production of malware that practically disappears out of sight of traditional signature-based detection systems. AI-enabled tools can study and adapt defense mechanisms, which is a great advantage for attackers as it helps them find a loophole in detection and persistence.In addition it can define real-time vulnerabilities and exploit them before patches can be released. Finally, GANs can be utilized to create realistic but fictitious data like fake network traffic or user credentials, which can easily cross the IDS as well as other security measures.

***Limits of Conventional Security***

The use of AI in cyberattacks brings forth many challenges to conventional security mechanisms Apruzzese et al. [70]. One of the main difficulties is that such attacks, driven by AI, adapt to the defense systems in a network, which makes it difficult for static rules or even signatures to detect them. In particular, such attacks with the use of AI adapt to the defense systems in a network, making it difficult for the static rules or even the signatures to detect them. Additionally, GANs and other AI techniques are capable of producing realistic but two-dimensional data to evade intrusion detection systems and other security measures. Furthermore, AI-enabled tools can identify and exploit vulnerabilities in real-time, allowing attackers to evade detection and maintain persistence within the networks.

***AI-Enabled Countermeasures***

In response to these challenges, counteraction through AI solutions becomes paramount when faced with the threats posed by AI-induced attacks Chowdhury et al. [116]. The counteracting measures employ the same advanced techniques that attackers use to detect and neutralize threats in real-time. These include AI-driven intrusion detection systems and adversarial ML, which aim to detect and mitigate such attacks effectively. Specifically, new AI techniques allow dynamic anomaly detection (Anomaly Detection (AD)) to detect and respond to attacks that are driven by AI—for example, ML algorithms can analyze network traffic patterns to discern abnormal activity that points toward potential threats. Moreover, training an AI model to detect and respond to attacks, known as adversarial ML, can counter threats such as those generated by GANs. Through techniques like competitive training and robust optimization, AI models can be fortified against malicious inputs. Additionally, AI models are capable of analyzing vast amounts of data in real-time to quickly identify and eliminate emerging threats, which is crucial for 6G networks due to their low-latency and high-reliability requirements.

***Decentralized AI and Distributed Ledger Technology (DLT) Security***

As far as decentralized AI systems are concerned, they operate without reliance on a single central authority, which aligns naturally with the distributed topology envisioned for 6G networks. Such systems distribute computation, data storage, and decision making across multiple nodes, supporting scalability, resilience, and low-latency operation [7], [21].

This architectural paradigm is well suited to 6G use cases such as ultra-reliable low-latency communications (URLLC), massive machine-type communications (mMTC), and enhanced mobile broadband (eMBB), where edge computing and distributed intelligence play a central role. Nodes in decentralized AI systems can act autonomously based on local data and interactions with neighboring nodes, enabling real-time decision making and efficient resource optimization [58].

Despite these advantages, decentralized AI systems face several challenges. Secure node-to-node communication is a primary concern, as distributed nodes must exchange data, model updates, or control information in environments that are vulnerable to eavesdropping, tampering, and man-in-the-middle attacks. Cryptographic techniques such as secure multi-party computation (SMPC) and homomorphic encryption are commonly proposed to protect data in transit [86]. Data leakage represents another critical risk, particularly when sensitive information may be unintentionally exposed during collaborative training or parameter exchange. To mitigate this issue, federated learning (FL) and differential privacy have been widely adopted. FL enables nodes to train models locally while sharing only model updates, whereas differential privacy introduces controlled noise to limit the disclosure of individual data records [87].

Faulty conditions and malicious node behavior further complicate decentralized AI deployments. Compromised or adversarial nodes may inject incorrect or misleading information, degrading model performance or destabilizing system operation. Byzantine fault tolerance (BFT) mechanisms and consensus algorithms, commonly employed in blockchain systems, provide resilience against such adversarial behavior by detecting and limiting the influence of malicious participants.

Blockchain and distributed ledger technologies (DLTs) have therefore been proposed as enabling infrastructures for secure and trustworthy decentralized AI. Their features include immutable record keeping, distributed consensus, and the use of smart contracts to automate and enforce system rules. In this context, blockchain can provide a transparent and tamper-evident ledger for recording data exchanges and model updates, while smart contracts facilitate verifiable coordination among distributed AI agents [117].

However, scalability remains a fundamental concern, particularly for large-scale deployments such as Internet of Things (IoT) and autonomous systems. Blockchain and DLT scalability is often constrained by consensus protocols and the overhead associated with maintaining a distributed ledger [118]. Energy consumption is another critical challenge, as consensus mechanisms such as Proof of Work (PoW) incur substantial computational and energy costs.

To address these limitations, alternative consensus mechanisms, including Proof of Stake (PoS) and Proof of Authority (PoA), have been proposed to improve energy efficiency and sustainability [119]. Nevertheless, their suitability for highly dynamic 6G environments requires further investigation.

Finally, regulatory, legal, and interoperability challenges remain largely unresolved. The deployment of decentralized AI systems raises concerns related to data protection, intellectual property, accountability, and liability. Moreover, the lack of standardized protocols and frameworks hampers interoperability and large-scale adoption. Establishing unified standards and governance mechanisms will be essential to enable the responsible and widespread deployment of decentralized AI in future 6G networks [2], [8], [23], [38], [39], [108].

***FL Threats and Protections***

Collaborative training across multiple devices is achieved using FL without acquiring raw data. Although it has become highly popular for privacy, it is susceptible to two types of attacks: model poisoning and inference attacks, in which the former adversely inject malicious updates and the latter reveal private information from model parameters. Thus, strong aggregation methods and safe communication protocols are needed for the secure use of FL in 6G [2], [5], [8], [21], [23], [29], [38], [39], [45], [118]

***Adversarial ML***

With adversarial ML, the inputs will be designed specifically with the intent of deceiving AI models. The impacts of an attack of this type would result in drastic disruptions in all 6G network functions and perhaps considerable degradation in service quality and frictions in security systems. Some of the defenses include adversarial training, input sanitization, and consistent modeling techniques Tang et al. [120]. For example, defensive distillation has been proposed to make models less sensitive to adversarial input [2], [3], [45].

***QoTE-Based Security Evaluation***

Another aspect is QoTE-based security evaluation. QoTE stands for Quality of Trust Experience and is an innovative metric introduced in evaluating trustworthiness in 6G AI systems. Factors such as reliability, transparency, and resilience of AI components are taken into account when establishing QoTE metrics. Despite challenges such as standardization and heterogeneous implementation contexts, it is useful in assessing and improving the security posture of AI-based 6G systems [4], [22].

***AI for Detection, Authentication, and Access Control***

AI could be one of the main elements in detecting and countering cyberattacks. With anomaly detection, behavior analysis, and predictive modeling, threats can be identified on the spot. However, the most important issue is to keep these AI systems impenetrable to evasion and poisoning attacks. Advanced systems have already shown a lot of promise in dynamic 6G environments, such as RL - based intrusion detection systems [1], [2], [23].

AI could also be trained to improve authentication and access control through linking different identified behaviors of a user, analysis of their biometrics and context parameters. Continuous authentication mechanisms, which verify a user during their whole session, greatly match the highly dynamic environments of 6G networks. Nevertheless, it is still a difficult challenge to find ways to make such systems resilient against spoofing attacks and keeping users' privacy [3], [4], [11].

***Blockchain for Secure Data and Transactions***

In addition, blockchain provides a decentralized and unalterable way for securing data and transactions in AI-powered 6G systems. This is a means of offering transparent and verifiable secure privacy protection mechanisms like identity management and consented data sharing between users. Scalability and latency are two major challenges when blockchain is integrated with a 6G network, which calls for new lightweight consensus algorithms and off-chain solutions [2], [4], [21].

***End-to-End Trust and Adversarial Readiness***

Furthermore, the end-to-end mechanisms of trust establishment will cover interaction across all levels of the 6G network. AI-enabled trust management systems will dynamically evaluate the trustworthiness of a user through real-time data. Creating a system that is protected against the manipulation of data and collusion is quite difficult. Under end-to-end trust mechanisms, secure communications are done across all layers constituting the 6G network. AI-driven trust management systems dynamically evaluate the trust of a user through real-time data. Unfortunately, it is quite a challenge to design the system so that it would be powerful enough against manipulation of data and collusion among participants [1], [39].

Additionally, adversarial learning refers to training the AI systems in performing reliably and providing an adversarial environment against different attacks. These would be really important for 6G networks, as they would face constant qualitative threats from the attackers. Furthermore, adaptive AI models evolve as the threats emerge and are a requirement. However, they need to be very stable in mechanism so that the models are prevented from being mocked while training and in deployment [3], [4], [29].

***Endogenous Security and Privacy at the Edge***

Last but not least, endogenous security from 6G intends to provide features of threat detection and mitigation at the source of the network itself. However, it remains very challenging to have powerful AI-based security solutions that can defend against adversarial attacks. Besides, the privacy issues of the distributed nature of 6G networks go beyond this. It will require quite advanced privacy-preserving techniques such as FL and secure multiparty computation to ensure the sensitive data is kept protected. At the same time, it is getting processed through multiple edge nodes. It would also need to strike a safety aspect with ubiquitous sensing capability that could be misused for unauthorized surveillance or data harvesting Li et al. [11].

### B. Efficiency and Sustainability Challenges

The integration of AI into 6G networks also brings forth substantial challenges related to efficiency and sustainability. While AI-driven optimization techniques have demonstrated their effectiveness in enhancing spectrum utilization, energy efficiency, and network throughput, they also introduce new concerns in terms of computational and environmental costs. The deployment of large-scale AI models and intelligent network components demands significant processing power and energy resources, potentially undermining the overall sustainability goals of 6G networks. Therefore, balancing high performance with low energy consumption and environmental impact remains one of the central challenges for future 6G infrastructure.

***Energy Efficiency in Asynchronous 6G Networks*** Asynchronous 6G wireless networks now demand unprecedented energy efficiency and sustainability requirements . With AI now at the center of 6G, energy consumption and environ-mental sustainability will be crucial in considering how these systems function. Depending on massive connectedness under ultra-low latency requirements, intelligent decision-making and further complicates energy requirements Saruthirathanaworakun et al. [37]. Therefore, ensuring seamless operation of the 6G network while pursuing sustainability is a multifaceted task [2], [3].

**TABLE 7. Efficiency and Sustainability Challenges and AI-Based Solutions in 6G Networks**

| Topic | Problem | AI-Based Solution |
|---|---|---|
| Energy Efficiency in Asynchronous 6G Networks | Massive connectivity and ultra-low latency requirements significantly increase energy consumption and sustainability complexity. | AI-driven intelligent decision-making to manage energy consumption under asynchronous 6G operation. |
| Sustainable Smart Environments | Pervasive sensing and massive IoT data generation demand high computational and communication resources. | AI-based data analytics and intelligent control enabling efficient operation of smart environments. |
| Smart Grids and Energy Management | Dynamic energy distribution and integration of renewable sources require real-time coordination. | AI-enabled predictive analytics and real-time energy management in smart grids. |
| Reducing the Energy Footprint of AI | Large-scale AI models consume significant computational power and energy. | Model compression, pruning, quantization, and green AI techniques for energy-efficient AI operation. |
| Renewable Energy and Wireless Power Transfer | Remote and distributed devices require sustainable energy supply. | Renewable energy harvesting and AI-assisted wireless power transfer for network components. |
| AI-Based Energy Optimization | Static energy allocation leads to inefficiencies and energy waste. | AI-powered dynamic energy allocation based on traffic patterns and resource utilization. |
| Collaborative Learning for Energy Efficiency | Centralized learning increases communication and computation energy costs. | Collaborative and decentralized learning reducing data transmission and computation overhead. |
| Optimizing Quality of Things' Experience (QoTE) | Energy efficiency must be balanced with service quality and user experience. | AI-driven QoTE optimization enabling energy-efficient resource allocation. |
| RL for Energy Efficiency | Manual energy management cannot adapt to dynamic network conditions. | DRL-based autonomous learning for energy-efficient resource allocation and management. |
| Mean-Field Game Theory for Resource Balance | Large-scale device interactions complicate balanced and energy-efficient resource usage. | Mean-field game models for energy-efficient resource balancing across massive networks. |
| Task Offloading and Wireless Energy Transmission | Resource-constrained devices face high energy consumption for local computation. | AI-optimized task offloading and wireless energy transmission to reduce device energy usage. |
| Collaborative Learning for Distributed Optimization | Redundant computation across devices increases energy consumption. | Collaborative learning for distributed optimization and improved energy efficiency. |

***Sustainable Smart Environments***

The development of future smart and environmentally friendly spaces is anticipated to enhance the quality of life for individuals by offering a variety of new, people-centered services. The integration of innovative applications can significantly benefit smart environments, including intelligent transportation, smart grids, and smart agriculture. However, these applications necessitate advanced capabilities in pervasive sensing, data mining, intelligent control, and distributed actuation systems, which present considerable challenges.

The substantial volume of data, generated by numerous IoT devices, can be effectively analyzed to provide timely responses, provided that high-quality wireless communication networks with robust computational capabilities are in place. Furthermore, the rapid growth of IoT devices calls for a considerable improvement in connection capacity and coverage for 6G technology Zhang et al. [32].

***Smart Grids and Energy Management***

To begin with, smart grids are expected to operate as the most critical parts in the energy management of 6G networks. These systems employ AI for predictive analytics and real-time decision making in managing the distribution of energy among components. Smart grids enable the integration of renewable energy sources with storage systems, such that they supply consistent power with little backup Hossain et al. [4].

***Reducing the Energy Footprint of AI***

Most of the AI algorithms used in 6G networks require a large amount of computational resources and, hence, consume a vast amount of energy. Currently, research is being conducted on topics such as model compression, pruning, and quantization that aim to reduce the energy footprint of AI operations while keeping the performance intact Yang et al. [2].

Moreover, green AI is a concept that encourages the value of building and using AI techniques with a minimum environmental footprint. This means creating algorithms that use less energy and promoting hardware designed specifi-cally for low-power AI processing, for example, low-power consuming and efficient hardware Du et al. [3].

***Renewable Energy and Wireless Power Transfer***

Methods of generating energy like solar, wind, or Radio Frequency (RF) energy collection will supplement the energy source of the network components. Wireless Power Transfer (WPT) is identified as the source of energy to be transferred to devices functioning remotely or outside but operating without their presence in daily use [2], [4].

***AI-Based Energy Optimization***

AI-powered power optimization algorithms dynamically allocate energy resources in accordance with the network demand and ensure that the overall efficiency is maximized. They adjust the energy consumption by understanding the utilization of the resources and traffic patterns so that the balance of energy can occur throughout the network Mir et al. [1].

Furthermore, dynamic energy allocation models deploy AI methods to predict and react to varying energy demands. Based on real-time data, these build up flexible power distribution to keep network performance at its best and at the same time avoid waste Hossain et al. [4].

***Collaborative Learning for Energy Efficiency***

At the same time, collaborative learning enables distributed devices to train shareable AI models without centralizing data while saving energy costs associated with data transmission and computation. It is an efficient and decentralized strategy that holds people's privacy intact [2], [3].

***Optimizing Quality of Things' Experience (QoTE)***

Optimizing the QoTE allows for the allocation of resources according to the wishes of users and the parameters of the network. AI algorithms come up with energy-efficient strategies along with offering a high-quality service Chi et al. [22].

***RL for Energy Efficiency***

In addition, DRL frameworks approach the problem of energy efficiency via autonomous learning of optimal policy for resource allocation and management of different networks. Dynamic environments will adjust to sustainable operations Yang et al. [2].

***Mean-Field Game Theory for Resource Balance***

Mean-field game theory mainly deals with modeling applications where there are a large number of devices interacting through a network. Using cognitive metrics, models can further improve the usage of balanced resources across the access network with energy-efficient ends [1], [4].

***Task Offloading and Wireless Energy Transmission***

Besides, task offloading is transferring computation tasks from a device to energy-saving nodes or even cloud servers, thus lessening the energy consumption of resource-constrained devices. AI helps optimize the offloading decision for reduced latency and power usage Du et al. [3].

Wireless energy transmission technologies also harmonize the recorded energy supply for IoT devices and their network components. AI improves the system's efficiency by redistributing energy and lowering the losses that occur Hossain et al. [4].

***Collaborative Learning for Distributed Optimization***

Lastly, collaborative learning is a type of learning in which various devices come together to improve the accuracy of AI models, as well as energy efficiency. This method is designed to reduce redundant computation while making effective use of distributed resources [1], [2].

### C. Hardware and Infrastructure Challenges

The realization of 6G networks heavily depends on advancements in hardware and infrastructure capable of supporting ultra-high data rates, low latency, and massive connectivity. While innovative hardware solutions such as reconfigurable intelligent surfaces (RIS), terahertz (THz) transceivers, and advanced antenna systems have shown great potential to enhance network performance, they also present significant design, deployment, and maintenance challenges. The integration of these emerging technologies requires substantial investment in new infrastructure and the development of energy-efficient, cost-effective components that can operate under demanding environmental conditions. Consequently, achieving scalable, reliable, and sustainable hardware and infrastructure solutions remains a major hurdle in the evolution toward fully realized 6G systems.

***AI-6G Hardware Overview***

AI-6G hardware includes pieces of equipment that are meant for the smooth operation and integration of AI into 6G networks. These include powerful processors, various accelerators such as GPUs (Graphics processing units), TPUs (Tensor Processing Units), and sometimes chips made for AI purposes, which interestingly enhance the delivery of data in a real-time decision-making system [99], [102].

An example is the use of neuromorphic processors that represent human neural systems, improving the efficiency and speed of AI computation manifolds. However, scalability and power reduction cost-performance trade-offs pose a challenge to achieve [1], [3], [4], [6].

***Quantum Computing and 6G Integration***

Quantum computing is nothing but a different method of proceeding computing, ruling leverages superposition and entanglement to perform computations that are impossible for classical computers. In the general context of quantum computing, 6G networks can enhance AI algorithms by introducing speed in optimization, cryptographic security, and complex problem-solving.

Seen from this perspective, it is useful in networks involving multiple massive deployments, in real-time data analytics, and in the enhancement of encryption protocols. Integration of quantum computing with 6G infrastructure still presents many challenges, including the development of quantum-compatible hardware, error correction techniques, and stable qubits that can function in real-world environments. However, if quantum computing and 6G networks

**TABLE 8. Efficiency and Sustainability Challenges and AI-Based Solutions in 6G Networks**

| Topic | Problem | AI-Based Solution |
|---|---|---|
| AI-6G Hardware Overview | High-performance AI processing requires powerful accelerators with challenging scalability and power-efficiency trade-offs. | AI-specific processors, accelerators, and neuromorphic hardware for efficient real-time AI computation. |
| Quantum Computing and 6G Integration | Quantum computing integration faces hardware compatibility, stability, and deployment challenges. | Hybrid quantum–classical architectures enhancing AI optimization, security, and analytics. |
| Quantum Computing Principles | Classical computing limitations restrict solving large-scale optimization and complex problems. | Quantum principles (superposition, entanglement, interference) enabling high-speed parallel computation. |
| Quantum Computing for 6G Applications | Complex optimization, analytics, and cryptographic challenges exceed classical capabilities. | Quantum algorithms for resource optimization, real-time analytics, and secure communications. |
| Quantum Integration Challenges | Quantum systems suffer from decoherence, noise, hardware instability, and integration complexity. | Quantum error correction and hybrid quantum–classical system designs. |
| THz Communication Hardware | Severe path loss, hardware complexity, and line-of-sight constraints at THz frequencies. | Advanced THz transceivers, signal modulators, and AI-assisted beamforming. |
| Edge Computing and AI Microchips | Real-time 6G applications require low-latency processing under strict energy constraints. | Energy-efficient edge AI microchips enabling local real-time decision making. |
| Low-Complexity and Energy-Efficient Designs | Resource-constrained devices face high energy and computational overhead. | Low-complexity hardware architectures and lightweight algorithms. |
| Hardware–Software Co-Design | Mismatch between AI algorithms and hardware limits performance and efficiency. | Joint hardware–software co-design optimized for AI workloads. |
| Massive MIMO and Adaptive Antennas | Large antenna arrays increase hardware complexity, cost, and energy consumption. | AI-assisted beamforming, adaptive antennas, and RIS-enabled hardware. |
| Distributed Server Architectures | Centralized processing increases latency and reduces fault tolerance. | Distributed server infrastructures with energy-efficient processing. |
| Programmable Networking Hardware (SDN/NFV) | Rigid hardware limits flexibility and dynamic resource management. | Programmable SDN/NFV-capable hardware enabling AI-driven control. |
| Liquid Software-Defined Architectures | Rapid network reconfiguration requires specialized high-speed hardware support. | FPGAs and high-speed memory enabling flexible software-defined operation. |
| Edge Orchestration and AI Management | Coordinating AI models across edge and core increases management complexity. | AI-enabled edge orchestration with dedicated accelerators. |
| THz Communication and Biological Effects | Potential long-term biological effects of THz exposure remain unclear. | AI-assisted analysis of biological data and exposure modeling. |
| AI for THz Safety and Optimization | THz performance optimization must be balanced with human exposure safety. | AI-driven spectrum management, beamforming, and safety-aware modeling. |
| EMF Exposure and Standards | Existing EMF standards may not fully account for dense 6G deployments. | AI-based monitoring and modeling of cumulative EMF exposure. |
| Regulatory Gaps and Collaboration | Lagging regulations and lack of harmonized safety policies. | AI-supported evidence generation for policy adaptation and regulation. |
| AI for Exposure Monitoring and Policy | Manual exposure assessment lacks resolution and scalability. | AI-based real-time exposure monitoring and policy-support simulations. |

are harmonized, a whole new era in AI applications and communications will open up [2], [4], [7].

***Quantum Computing Principles***

Quantum computing operates on an entirely different set of principles from classical computing. Some of the main principles are: First, qubits can exist in a superposition of both 0 and 1 states, whereas classical bits can only be in one state at a time, enabling quantum computers to tackle a very large number of possibilities simultaneously.

Furthermore, qubits can be entangled, meaning the state of one qubit is dependent on the state of another even when separated by distance, allowing quantum computers to solve complex problems with comparatively less computation. Finally, quantum algorithms rely on interference to reinforce correct solutions while canceling wrong ones, thereby achieving higher speed and accuracy in computations.

***Quantum Computing for 6G Applications***

Quantum computing can completely change the perspective in which 6G works by enhancing AI algorithms, thus improving network performance and security from a cryptographic point of view. Specifically, quantum computing can solve complex optimization problems like resource allocation, route finding, and spectrum management much more efficiently than classical algorithms—for instance, the Quantum Approximate Optimization Algorithm (QAOA) can optimize base station locations in 6G networks for maximum coverage and minimum interference Farhi et al. [103].

Additionally, it enables real-time data analytics, processing vast amounts of data instantaneously for predictive maintenance, anomaly detection, and traffic prediction; quantum ML algorithms, for example, can analyze network traffic patterns to detect and neutralize cyberattacks Shor [121]. Notably, while quantum computing poses threats to cryptographic security, it also offers solutions like quantum key distribution (Quantum Key Distribution (QKD)) to establish ultra-secure 6G communications Schuld et al. [91]. Finally, quantum acceleration of AI and ML—exemplified by quantum neural networks processing complex datasets more efficiently than classical counterparts—further underscores its transformative potential for 6G systems.

***Quantum Integration Challenges***

Despite the far-reaching potential of quantum computing, several challenges surface in its integration into 6G networks. Foremost among these is the development of quantum-compatible hardware, as current quantum computers require extremely low temperatures and remain highly sensitive to ambient disturbances Preskill [101].

Equally critical is the issue of error correction, since quantum systems are prone to decoherence and noise, necessitating robust quantum error-correcting codes to ensure reliable operation. Another major hurdle lies in achieving stable qubits, as most existing qubits exhibit poor stability and depend on complex control systems. Finally, seamless integration with classical systems poses its own challenges, requiring innovative architectures and protocols to develop hybrid quantum-classical algorithms that can leverage the strengths of both paradigms Perdomo et al. [92].

***THz Communication Hardware***

THz communication is among the key aspects of the 6G infrastructure for high-end data rates since this new paradigm works between 0.1 and 10 THz frequency spectra. It can be used for extremely high-speed data transfer in an extensive range with applications, such as real-time holography and extended reality, which are bandwidth-demanding ones.

However, THz communication is required to address certain issues, including reduction of the signal itself, the need for a lot of novel hardware such as miniature THz transceivers and efficient signal modulators, to maintain the line of sight with reliable connectivity [1], [22], [38].

***Edge Computing and AI Microchips***

On the other hand, edge computing makes latency reduction and increases efficiencies by processing data closer to its origin. It is important, especially for 6G, where applications needing real-time capabilities include autonomous vehicles and industrial automation and will require instantaneous decision-making.

Edge AI-specific microchip features such as energy efficiency combined with high computational power are critical to this approach. Unfortunately, designing chips requiring good performance along with power efficiency is a major problem that must be solved before these technologies can be widely deployed [4], [29], [45].

***Low-Complexity and Energy-Efficient Designs***

With all its simplicity, complex low designs are responsible for the minimum energy consumption and computational overhead in 6G hardware. These low-complexity designs rely on simple algorithms and hardware architectures that yield performance with less resource burden.

This is particularly critical to devices under resource-constrained environments such as IoT sensors deployed in hollow grounds where energy for the environment is very limited [2], [21].

***Hardware-Software Co-Design***

In addition, hardware-software co-design paves the way for the incorporation of AI in 6G systems through co-design of hardware-software components. High-performance models can be created through effective co-design methods that specifically suit features like memory and processing units to the requirements of the AI algorithms.

Co-design further aids adaptive and efficient resource management, especially in cases of dynamic scenarios such as those that 6G can support with multiple variants of applications [4], [39].

***Massive MIMO and Adaptive Antennas***

Massive MIMO (MIMO) underpins the capacity and coverage advancements in 6G networks through the deployment of antenna arrays of huge scales. Translated into the available hardware along with specializations related to low-power amplifiers, beamforming modules, and high-precision RF transceivers, such advancements become critical for scal-ing massive MIMO systems toward cost-effective, energy-efficient high-performance operation [1], [8].

6G adaptive antenna technologies are also required for dynamic beamforming and user tracking. Such antennas tend to vary their parameters in real time and hence channelize maximum signal strength while minimizing interference. Innovations in materials like reconfigurable intelligent surfaces (RIS), hardware forms for adaptive control mechanisms, etc., are crucial to facilitating recently efficient and flexible communications in numerous 6G-based deployment cases [2], [22].

***Distributed Server Architectures***

A distributed server architecture is destined to become the backbone for processing enormous data volumes in the future 6G networks. Geographically dispersed servers decentralize data processing, thus reducing latency and enhancing fault tolerance.

Such a server architecture totally relies on innovations in server hardware-capacity storage and energy-efficient processors to make this infrastructure scalable and reliable [4], [38].

***Programmable Networking Hardware (SDN/NFV)***

The flexibility and programmability that are indispensable for 6G networks are achieved by integrating software-defined networking with network function virtualization. The hardware decouples network control functions, enabling dynamic resource allocation and efficient traffic management.

Hardware capable of supporting SDN/NFV, such as programmable switches and network processors, is vital for realizing the potential of AI-driven 6G systemsSaruthirathanaworakun et al. [37].

***Liquid Software-Defined Architectures***

Dynamic network services are delivered by liquid software-defined architectures with adaptive, rapid, fluid, and versatile software frameworks responding to the ever-changing dynamics in the network.

However, the key hardware requirements for high-speed reconfiguration and rapid update capabilities include field-programmable gate arrays (FPGAs) and high-speed memories. Such developments are intended to modernize modern 6G systems for flexibility and responsiveness to changing user demands Du et al. [3].

***Edge Orchestration and AI Management***

Finally, the unique edge orchestrators created for AI will be tasked with managing how AI models are deployed and managed at the network edge. They will require hardware that will perform distributed computing tasks, real-time data analytics, and seamless communication with central systems.

The breakthrough development of the edge accelerator along with AI-optimized networking equipment will pave the road for enabling such intelligent orchestration in a variety of applications in 6G [1], [7].

***THz Communication and Biological Effects***

The takeoff of terahertz (THz) frequency communication systems is one of the critical innovations that the increasingly advancing 6G technologies would involve. However, it opens a new domain of ongoing discussions regarding the health effects of using THz radiation for a long time. Operating at the frequency range of 0.1–10 THz, THz rays increase photon energy levels higher than those of microwave irradiation but below those associated with ionizing radiation. Although they do not ionize atoms or molecules, their effects on biological tissues remain largely unexplained. Initial research studies suggest that there might be some effect brought about by THz-wave propagation under such conditions, for example, some changes at the cellular level concerning protein folding and DNA integrity. Thus, there is an urgent call for more comprehensive research to assess the long-term exposure risks and define safety thresholds. As 6G intends to embed the daily frequencies of THz into applications, identifying and clearing issues beforehand will be critical for public health and safety through rigorous scientific inquiry [3], [21], [23].

***AI for THz Safety and Optimization***

AI is not only a key task maker but also an advisor in the process. On the one hand, the performance, management, or minor improvements of TeraHertz (THz) communication systems by real-time spectrum sharing, beamforming, and adaptive signal processing are quite interesting with the occasion of AI intervention. On the other, AI may be used to speed up biomedical research by analyzing vast sets of biological data to identify subtle changes posed to cellular or molecular structures due to THz exposure. AI simulations and predictive models also render the development of a safer environment for THz-powered devices and infrastructure, standing guard against human exposure. With such new systems that endeavor to flock to everyday applications with the use of THz frequencies by 6G, the inclusion of AI is paramount in guiding the responsible development and health-consciousness of such transformational technology.

***EMF Exposure and Standards***

Higher density deployments of stations, devices, and sensors are also expected as the predicted increase in the development of 6G networks emerges. It would therefore follow that the emissions will become denser and create a more increased exposure to electromagnetic fields . Concern has been raised on whether the existing EMF standards are adequate to care for health impacts from 6G technologies. Most current standards, for example, that of the International Commission on Non-Ionizing Radiation Protection (ICNIRP), were set for lower frequency ranges and do usually not incorporate the uniqueness of their characteristic bands, such as THz frequency. Furthermore, cumulative effects resulting from a gradual immersion through low radiation levels in urban settings are still poorly understood [1], [4], [38].

***Regulatory Gaps and Collaboration***

Since there is a considerable increase in consensus among scientist communities about health risks, the health regulators have increasingly fallen behind in their principles, updating safety standards, thus filling in the gaps that will not risk public health in the 6G launching. New and stronger international collaboration and adaptable policy frameworks would play significant roles in addressing these risks while making possible safe deployments of advanced wireless technologies in the future.

***AI for Exposure Monitoring and Policy***

In this context, AI is being hailed as one of the most important emerging mechanisms for proactive risk and regulatory action and adaptation. Through AI, real-time monitoring and modeling of EMF exposure across urban space could provide high-resolution information at the scale of cumulative exposure. The same can go a long way toward linking health outcome indicators to environmental EMF level statistics for most epidemiological studies, making possible a more evidence-based threshold determination for safety. Further, AI simulations could support policymakers' efforts to alert the community to possible future health effects before debate regarding changes occurs. With growing scientific consensus about possible health risks, bringing research and regulation together will demand stronger international collaboration.

### D. Standardization and Ethics

The development of 6G networks introduces complex challenges in terms of standardization and ethics, as global collaboration is required to ensure interoperability, fairness, and responsible deployment. Unlike previous generations, 6G is expected to integrate diverse technologies such as AI, blockchain, quantum communication, and intelligent sensing, which demand unified frameworks and regulatory agreements across nations and industries. The absence of well-defined global standards may lead to fragmentation, security vulnerabilities, and inequitable access to advanced communication services. Moreover, ethical concerns surrounding data ownership, algorithmic transparency, and digital inclusion must be addressed to prevent biases and misuse of AI-driven decision-making in 6G ecosystems. Therefore, establishing robust international standards and ethical guidelines is critical to ensuring that 6G technologies evolve in a trustworthy, inclusive, and globally harmonized manner.

***Standardized AI Protocols and Interoperability***

With standardized AI protocols for 6G networks , the efficiency of cooperation and the assurance of security are guaranteed. Current challenges are defining and ensuring the commonality of format for data exchange, interoperability for all AI models, and the existing infrastructure of the network. These standards of AI also are significant in ensuring the attainment of ethical problems relating to their transparency and accountability in AI decisions [22], [23]. Above all, changes in the regulatory frame rules have to be necessitated as a result of the 6G networks to address issues like the protection of data, responsibility in AI, and cross-boundary data flow, such as innovation, and individual interests in regulation and policy. AI-based decision-making creates complexity concerning the legal and ethical accountability of such decision-making action, hence requiring powerful international collaborations. For instance, the GDPR set up in the European Union is a precedent regarding harmonized data protection standards extendable also to 6G systems [39], [45].

Again, interoperability involving devices and platforms will be a significant challenge for 6G networks when they introduce among themselves various AI-enabled applications and edge devices. Such incompatibility may not allow easy data exchange and collaborative AI operations. Standardized APIs and protocols become critical in disassembling such barriers. Distributed ML models and network slicing compatibility in 6G are critical to ensure efficient resource use and system performanceSaruthirathanaworakun et al. [37].

***QoTE and Ethical Considerations***

Furthermore, the QoTE encompasses ethical considerations in terms of fairness, equity, and welfare and exceeds both QoS and QoE. When such systems are enabled in the 6G environment and handle hundreds of thousands of IoT devices, the equality of access and the non-discriminatory nature of treatment will come into the picture.

AI models have to interpret ethical rules to help make system behavior consistent with societal values. These technologies will adapt QoTEs dynamically to keep users optimally happy, and within ethical standards. Real-time analytic capabilities can be harnessed to ensure that there are undetected biases within the service provision. The AI-enabled QoTE systems will also ensure that users are confident of network behaviors and decisions, thus encouraging a more diverse digital ecosystem Chi et al. [22].

***Explainable AI (XAI)***

Explainable AI (XAI) will also open the black boxes of AI-based complex systems within 6G networks. AI penetrates maturing critical applications such as healthcare and autonomous cars, particularly when the decision process has to be recognized and responsible. Advanced techniques in XAI, such as model-agnostic explainability and interpretable neural networks, underscore the importance of transparency in the deployment of AI in this context. Moreover, directed regulations such as the GDPR "right to explanation" highlight the relevance of XAI in compliance as well as trust by the user Wang et al. [39].

***Regulatory Compliance and Accountability***

The involvement of AI in 6G networks must conform to regulatory frameworks that stress transparency and accountability. For example, there is the GDPR in the European Union that consists of a 'right to explanation' meaning that individuals must be given information that is meaningful to them on decisions taken using automated systems. Thus, the regulation emphasizes XAI integrity in compliance and building user trust Wachter et al. [35].

**TABLE 9. Standardization and Ethics Challenges and AI-Based Solutions in 6G Networks**

| Topic | Problem | AI-Based Solution |
|---|---|---|
| Standardized AI Protocols and Interoperability | Lack of common AI protocols, data formats, and interoperability across heterogeneous 6G devices and platforms hinders efficient cooperation and secure AI operation. | Standardized AI protocols, common data exchange formats, and interoperable APIs enabling collaborative AI operation and secure integration across 6G infrastructures. |
| Regulatory Frameworks and Accountability | AI-driven decision making in 6G complicates legal responsibility, data protection, and cross-border data governance. | AI-aligned regulatory frameworks and policy mechanisms ensuring transparency, accountability, and compliance with data protection regulations (e.g., GDPR). |
| QoTE and Ethical Considerations | AI-enabled 6G systems risk unfairness, bias, and unequal access when managing massive numbers of users and IoT devices. | AI-driven QoTE mechanisms dynamically optimizing fairness, inclusiveness, and ethical service provisioning in large-scale 6G environments. |
| Explainable AI (XAI) | Black-box AI models reduce transparency and trust, particularly in critical 6G applications such as healthcare and autonomous systems. | Explainable AI techniques providing interpretable models and transparent decision-making to enhance trust and regulatory compliance. |
| Regulatory Compliance and Transparency | Automated AI decisions must satisfy legal requirements such as the right to explanation and ethical accountability. | XAI-enabled AI systems supporting auditable, transparent decisions aligned with regulatory and ethical standards. |

## E. Communication and Networking

The evolution toward 6G networks presents profound challenges in communication and networking design, driven by the demand for ultra-reliable, low-latency, and high-capacity connectivity. As network architectures become increasingly complex—integrating heterogeneous access technologies, intelligent routing, and dynamic spectrum sharing—ensuring seamless communication across diverse devices and environments remains a significant technical obstacle. Furthermore, achieving efficient coordination among terrestrial, aerial, and satellite networks introduces additional requirements for synchronization, interoperability, and adaptive resource management. The incorporation of AI into network control and optimization, while promising enhanced performance, also raises concerns regarding scalability, transparency, and fault tolerance. Hence, addressing these communication and networking challenges is essential for realizing the full potential of 6G systems in delivering ubiquitous and intelligent connectivity.

### 1) 6G Communication Service Types

***mMTC***

Firstly, the Massive Machine-Type Communications (mMTC) in 6G will allow for an unprecedented density of device interconnection that could reach millions per square kilometer. This is necessary for smart cities, industrial IoT, and environmental monitoring. mMTC faces challenges, such as energy efficiency, scalability, and interference management. AI-based solutions, including FL and edge computing, optimize resources and offer intelligent coordination of the devices. The ability to integrate mMTC with other 6G technologies, such as ultra-reliable low-latency communication (URLLC), will further augment mMTC's features to offer uninterrupted connectivity for several use cases [2]–[5], [7], [21], [26], [29], [38], [39], [45].

mMTC represents one of the vital aspects of today's 6G networks, enabling massive scaling for the great number of IoT devices deployed across various verticals. The key characteristics of mMTC in 6G include: High device density, designed to connect millions of devices per square kilometer, particularly crucial for smart cities where thousands of sensors must communicate simultaneously. Low power consumption is equally critical, as most battery-powered IoT devices require energy-efficient protocols; mMTC incorporates technologies like LPWAN to minimize energy use and extend operational lifetime. Scalability stands as another essential feature, with mMTC needing to accommodate the projected tens of billions of IoT devices by 2030, demanding novel solutions for network resource management. Finally, heterogeneous connectivity allows mMTC to serve diverse devices ranging from low-data-rate sensors to high-bandwidth industrial machines Chowdhury et al. [116].

mMTC will find applications in many fields. For a start, Smart Cities, where it will support deployment of common infrastructure like traffic management systems, environmental monitoring sensors, and smart grids, with thousands of sensors monitoring air quality, traffic flow, and energy consumption in real-time. Industrial IoT (Industrial IoT (IIoT)) will utilize mMTC to connect machines, robots, and sensors for monitoring manufacturing processes, equipment health, and predictive maintenance. Environmental Monitoring applications will leverage mMTC to deploy sensors in remote and harsh environments for tracking climatic conditions, wildlife, and natural disasters, such as forest sensors for early wildfire detection or ocean sensors for water quality monitoring. Agriculture will benefit through precision farming enabled by mMTC-connected sensors and actuators that monitor soil conditions, weather patterns, and crop health, helping farmers optimize irrigation, fertilizer application, and pest control Gubbi et al. [122].

mMTC has the promise of supporting a range of services needing very low throughput and latency over wireless technologies, yet several challenges must be overcome to fully realize these capabilities in 6G networks. Energy efficiency remains critical as most IoT devices run on battery power, requiring energy-efficient communication protocols to extend operational lifetime - necessitating integration of low-power wide-area network technologies into mMTC for 6G. Scalability presents another key challenge, with mMTC needing to support billions of IoT devices by 2030, driving the need for novel approaches to network resource management and efficient communication. Interference management becomes crucial due to the high device density in mMTC networks, demanding advanced techniques like beamforming and cognitive radio to maintain network performance. Finally, heterogeneous connectivity requirements mean mMTC must accommodate a wide spectrum of devices with varying communication parameters, requiring highly flexible and adaptable communication protocols.

AI-based solutions attempt to counter the next challenging path for mMTC in 6G networks through several key approaches. FL enables distributed devices to cooperatively train an AI model without sharing raw data, ensuring privacy while minimizing communication overhead - particularly valuable for mMTC's resource-constrained devices with limited computation and energy resources. Edge Computing reduces latency and alleviates centralized cloud burdens by processing data closer to its source, enhancing real-time applications like industrial automation and environmental monitoring. Dynamic Resource Allocation employs AI algorithms to optimize network resources (bandwidth, power) based on real-time conditions, such as prioritizing emergency sensors during network congestion. Finally, Interference Mitigation leverages AI techniques like DRL to optimize spectrum usage in mMTC environments, where DRL dynamically allocates spectrum resources to minimize interference and improve overall network performance.

Integrating mMTC with other 6G technologies such as ultra-reliable low-latency communication (URLLC) and integrated sensing and communications (ISAC) will enhance mMTC capabilities and enable new use cases. URLLC integration with mMTC creates potential for mission-critical applications like remote surgery and autonomous vehicles that require both massive connectivity and ultra-low latency characteristics. ISAC integration combines sensing and communication in a single system, enabling applications such as environmental monitoring and obstacle detection; when merged with mMTC, this allows real-time information collection and communication in dynamic environments Shi et al. [49].

***eMBB***

Enhanced Mobile Broadband (eMBB) constitutes a fundamental service class in 6G networks, enabling ultra-high data rates reaching up to 1 Tbps and significantly improved spectral efficiency. These capabilities support bandwidth-intensive applications such as holographic communication, extended reality (XR), and ultra-high-definition video streaming. However, achieving consistent performance under high mobility and increasing power demands introduces major challenges. AI-enabled techniques, including adaptive modulation, advanced channel coding, intelligent beamforming, and AI-driven network slicing, are essential to ensure reliable and efficient eMBB service provisioning in 6G environments [2], [3], [5], [7], [8], [21], [23], [26], [38], [45].

Key characteristics of eMBB in 6G include ultra-high data rates, enhanced spectral efficiency through advanced multiple access and modulation schemes, low-latency support for real-time applications, and robust performance under high-mobility conditions such as high-speed transportation and autonomous systems. These features position eMBB as the backbone of next-generation immersive services, including real-time holography, XR applications, and ultra-high-resolution video streaming Chowdhury et al. [116].

Despite its potential, eMBB faces several challenges in 6G networks. High power consumption arises from the computational complexity required to sustain extreme data rates, necessitating energy-efficient hardware and algorithms. Supporting high mobility demands adaptive signal processing techniques capable of maintaining link reliability, while coexistence with URLLC and mMTC services requires intelligent network slicing and dynamic resource allocation Mir et al. [1].

AI-based solutions address these challenges through adaptive modulation and coding schemes that dynamically adjust transmission parameters based on channel conditions, AI-enhanced beamforming for precise signal focusing in dense environments, and predictive analytics for traffic-aware resource optimization. AI-driven network slicing further enables tailored bandwidth and latency provisioning for eMBB services, ensuring consistent performance even during peak network load Shi et al. [49].

Finally, the integration of eMBB with other 6G technologies further extends its applicability. Combining eMBB with ultra-reliable low-latency communication (URLLC) enables mission-critical applications requiring both high data rates and stringent latency guarantees, while integration with integrated sensing and communication (ISAC) supports real-time data collection and communication in dynamic and sensing-intensive environments.

***URLLC***

Ultra-Reliable Low-Latency Communication (URLLC) in 6G networks enables mission-critical applications such as autonomous vehicles, remote surgery, and industrial automation, requiring end-to-end latency as low as 0.1 ms and reliability up to 99.9999%. These stringent requirements drive a paradigm shift in network architecture, incorporating advances in error correction, real-time processing, and distributed intelligence. AI-based predictive maintenance and fault detection enhance system reliability, while edge computing and distributed AI significantly reduce latency. Furthermore, the integration of URLLC with other 6G technologies, such as integrated sensing and communication (ISAC), expands its functional versatility [2]–[7], [21], [23], [26], [29], [38], [39], [45].

The technical foundations of URLLC in 6G build upon 5G enhancements and extend them to meet next-generation application demands. Ultra-low latency is achieved through innovations in radio access networks, network slicing, edge computing, shortened transmission time intervals, and preemptive scheduling. High reliability is ensured via advanced channel coding, hybrid automatic repeat request (HARQ), and diversity schemes, enabling robust fault tolerance and consistent communication Popovski et al. [123]. Edge computing further supports early data processing close to end devices, while edge AI enhances real-time decision-making for latency-sensitive URLLC services Mao et al. [99].

AI-driven enhancements play a central role in URLLC performance optimization. Predictive maintenance leverages AI to anticipate equipment failures and network anomalies, enabling proactive intervention and reduced downtime. Real-time fault detection and mitigation mechanisms allow automatic traffic rerouting and resilience against failures, maintaining uninterrupted URLLC operation. Edge AI deployment enables localized, low-latency decision making essential for mission-critical scenarios such as autonomous systems and industrial control.

Finally, interoperability with other 6G technologies further strengthens URLLC capabilities. Integration with ISAC combines sensing and communication within a unified framework, supporting applications such as environmental monitoring, obstacle detection, and gesture recognition. The addition of URLLC-grade reliability and latency guarantees to ISAC significantly broadens the scope and robustness of real-time 6G applications Liu et al. [30].

### 2) AI-Enabled Networking and Control Mechanisms

***Spectrum Sharing***

The need for spectrum sharing under 6G will become an inevitable requirement to offer any new bandwidth to an already high demand. This will be done with Digital Signature Algorithms and cognitive radio technologies for finding and exploiting underutilized frequency bands Liu et al. [30]. However, interference mitigation, regulatory compliance, and coordination of multiple stakeholders will remain challenges. AI-based spectrum sensing and allocation algorithms can optimize real-time spectrum usage, thereby ensuring proper and efficient sharing across different applications and users [3], [7], [29], [37], [39], [45].

***ISAC***

Also, ISAC as a step into the world of the 6G shall present complementary functionalities of communication and sensing. Applications range anywhere from environmental monitoring through gesture recognition to collision avoidance. Such advanced multifunctional offering using ISAC shall enable sensing with communication and vice versa, thus allowing better resource usage. The circle of problems involved entails avoidance of interference from sensing and communication signals and achieving proper sensing under dynamic environmental conditions. AI algorithms such as deep learning applied for signal processing and pattern recognition would play a vital role in optimizing the performance of the ISAC [29], [38], [45].

***Spectrum Monitoring and Localization***

Spectrum monitoring and localized flourishes into 6G for tracking the spectrum usage in real-time and locating the device to the last meter. It is a requirement for the timely response to an emergency, asset tracking, and network-level optimization. Nevertheless, some challenges are being able to maintain accuracy in varying complex scenarios, and possibly protection and security. An AI-based approach with FL and RL techniques can improve spectrum monitoring and localization by providing a framework for distributed and adaptive decision-making [29], [37], [38].

***Border Spectrum Coordination***

Border spectrum coordination will help 6G deal with the challenges of spectrum use management on a national as well as regional scale. This is important for applications like satellite communications and roaming across countries. With AI-based negotiation and coordination mechanisms, spectrum sharing can be implemented smoothly and conflict management can be realized, thus ensuring timely and fair resource utilizationSaruthirathanaworakun et al. [37].

***Dynamic Routing***

Dynamic routing in 6G will support that adaptive and rapid data transfer carried on heterogeneous networks. This is very much necessary due to the increasing complexity and variability of network traffic. These include managing scalability, minimizing latency, and reliability issues. AI-based routing algorithms like multi-agent DRL can help optimize routing decisions in real time for better performance and resilience of the network Tshakwanda et al. [6].

***Self-Healing Networks***

Self-healing networks in 6G are also required to detect and heal themselves from faults without any impact on the service. This is paramount to obtaining reliable large-scale and dynamic networks. AI-based mechanisms for fault detection, cause analysis, and self-healing are aimed at achieving reduced downtime and operational costs through automation Chataut et al. [7].

***Virtual Network Slicing***

AI will enhance virtual network slicing , enabling the efficient realization of specific applications and users. This requires dynamic resource allocation, real-time optimization, and efficient management of network slices. AI algorithms such as deep learning and RL can optimize slice throughput and thereby contribute to quality of service (QoS) and resource efficiency [1], [4]–[7], [26], [38], [39], [45].

***Intelligent Network Slicing***

The use of intelligent network slicing by technologies such as SDN and NFV will ensure network flexibility and scalability in 6G networks. AI-controlled orchestration and control of the SDN and NFV resources will ensure network slice deployment and operation, which can support distinct use cases with different requirements [2], [5], [7], [8], [22], [23], [39], [45].

***Grant-Free Schemes***

Grant-free schemes will allow devices to transmit data without any prior scheduling. This is particularly beneficial to mMTC and URLLC applications. There are challenges to understand and resolve collisions and to ensure reliability. The movement of AI-based collision detection and resolution algorithms will enable grant-free designs to operate with enhanced performance and provide effective and reliable communications Hossain et al. [4].

***Decentralized ML***

Decentralized ML within 6G will result in network-wide distributed intelligence to reduce latency and enhance scalability. It plays a crucial role in applications such as FL and edge AI. The challenges that need to be addressed in achieving these include data privacy, synchronization, and model accuracy. Decentralized ML frameworks can be adopted to solve these challenges, including the use of blockchain-based solutions, thereby making learning secure and distributed efficiently Hossain et al. [4].

***Distributed QoTE***

Distributed QoTE in 6G will manage user experiences across applications and devices everywhere with uniformity and high quality. This involves real-time monitoring and analysis of all network performance-related parameters to ensure optimization. The QoTE management system based on AI would dynamically revise the network parameters to provide a suitable user experience Chi et al. [22]. QoTE-based prioritization in 6G will facilitate network resource allocation on the user experience metric to allow critical applications to get the resources they need. AI prioritization algorithms will allow for real-time optimization of resource allocation to help with network utilization, resource maximization, and increased user satisfaction Chi et al. [22].

***Intelligent Routing and Traffic Forecasting***

In 6G, AI-driven intelligent routing and traffic forecasting will help optimize data transfers and mitigate possible congestion. ML algorithms can analyze historical traffic patterns and predict current demands so that routing decisions can be taken in advance. This will further support and enhance the performance and reliability of the networks under high-load conditions [3], [6], [38].

***Multi-Agent DRL***

Multi-agent DRL in 6G will support the collaborative decision-making process among multiple entities in the network to allocate resources efficiently and coordinate activities. This is crucial for a dynamic and complex environment. Multi-agent DRL can optimize the performance of the network to operate efficiently and adaptively Cao et al. [29].

***Dynamic Resource Pricing***

In addition, dynamic resource pricing in 6G will permit flexible and fair allocation of network resources based on demand and availability. AI-assisted pricing models can adjust prices dynamically and optimally for resource utilization and revenue generation Yang et al. [2].

***Tenant-Specific Network Virtualization***

Tenant-specific network virtualization in 6G will also facilitate dedicated custom virtual networks for distinct users and applications. AI-centric orchestration and management will ensure efficient resource allocation and optimization of performance as per the specific requirements of each tenant Chi et al. [22].

## VII. Applications and Use Cases

Applications for 6G include autonomous systems, such as self-driving cars, drones, and robotic systems (Fig. **??**). These will require ultra-reliable, low-latency communication to make real-time decisions based on detecting obstacles and route optimization. With AI embedded in 6G, such systems will not only detect but also predict environmental changes and self-adjust, enabling a quantum leap in safety and efficiency. For example, AI-driven channel estimation and spectrum management can optimize the performance of ultrabroadband technologies like terahertz communication, which is vital for seamless operation in autonomous systems [3], [23].

The integration of AR and VR with 6G would represent a highly immersive and contextually aware experience. Virtual meetings, remote training, and entertainment are examples of applications enabled by high bandwidth and ultra-low latency in 6G networks. AI-powered real-time data processing will enhance these applications with unprecedented personalization of experiences and efficiency in resource management. Distributed edge intelligence will be key in ensuring smooth operation of latency-sensitive AR/VR applications, especially under dynamic network conditions [2], [45].

The convergence of 6G with AI will benefit the healthcare domain in multiple ways. Low latency and edge computing will enable real-time applications such as remote patient monitoring, robotic-assisted surgery, and real-time diagnostics. AI algorithms can analyze data derived from wearables to predict health anomalies, thus enabling preventive care. Privacy concerns will also be addressed through localized data processing, ensuring sensitive patient information remains secure [4], [38].

Smart cities are among the most important use cases for 6G, as they integrate AI-driven solutions for efficient urban infrastructure management. For instance, distributed ML models will process data from IoT devices to monitor traffic, optimize energy consumption, and improve public safety. Dynamic network reconfiguration enabled by 6G will ensure that the vast number of connected devices in a smart city work seamlessly—even during congestion or emergencies [5], [22].

Additionally, transformational change in industrial automation will be the trademark of 6G. AI-based predictive maintenance will result in zero-downtime machinery, which means a predictive failure mode analysis from the sensor data itself. The smart offloading will make edge computing more computationally efficient to guarantee real-time decisions in complex industrial environments, resulting in adaptive, efficient, and sustainable processes [1], [3]. 6G will provide remote education in a more interactive and inclusive environment. AI-empowered platforms offer personalized learning, real-time translation, and wider access to learners around the world. Collaboration tools will be virtual classrooms and holographic meetings, powered by 6G's ultra-high speed and low latency for immersive and interactive environments for learners and professionals Chataut et al. [7]. Moreover, holographic communications are considered a key pillar for 6G, with live 3D interactions for every industry: medical, education, and entertainment. AI algorithms can effectively perform compression, transmission, and reconstruction in holographic communication with low latency, ensuring high-level visualization quality. With this, long-distance collaborations would feel just the same as it would face-to-face [7], [39]. Additionally, holographic RF systems based on Large Intelligent Surface (LIS) (large intelligent surfaces) enable enhanced control over the electromagnetic environment via spectral holography and spatial wave synthesis, resulting in improved spectrum efficiency and network capacity. However, several challenges remain to be addressed, including the optimal deployment of passive reflectors and metasurfaces, passive information transfer, channel state information (CSI) acquisition, and the need for low-complexity designs, all of which are essential for realizing the full potential of holographic radio technology. The rapid advancements in antenna technologies, including metasurfaces and reconfigurable intelligent surfaces (RISs), sometimes referred to as large intelligent surfaces (LIS), have garnered increasing interest in recent years. Unlike traditional antenna arrays, LIS can surpass the limitations associated with half-wavelength designs while maintaining low costs and energy consumption. This is achieved through a spatially continuous radio signal transmission and reception aperture created by an intelligent surface composed of numerous passive reflecting elements that can be controlled in terms of phase or amplitude. By effectively manipulating the radio propagation environment, RIS-based intelligent radio systems have the potential to overcome challenging propagation conditions, particularly in the context of 6G wireless communications. Furthermore, the development of holographic radio technologies, such as holographic MIMO and holographic beamforming (HBF), is progressing rapidly, facilitated by LIS. Holographic MIMO allows for the shaping of electromagnetic waves towards specific targets using an economical transformative wireless planar structure made up of sub-wavelength metallic particles. HBF, a novel beamforming technique that utilizes software-defined antennas (SDAs), employs a dynamic beamforming architecture based on passive electronically steered antennas. Through holographic recording and reconstruction, HBF can generate desired beams flexibly, achieving spatial resolutions that exceed those of conventional beamforming techniques Zhang et al. [32].

AI will play a crucial role in the allocation of resources for 6G networks, ensuring optimal bandwidth, computing power, and energy consumption. Real-time analytics will enable the dynamical change in resource distribution according to applications' diverse requirements, such as AR/VR, healthcare, and IoT. These systems will ensure efficient utilization of the network while maintaining quality of service [6], [37] Distributed ML where devices train models together while respecting privacy of the data of others characterizes 6G. In a decentralized manner this will help greatly lower dependence on centralized cloud architecture and lean towards faster and more efficient learning. Distributed ML will improve the scalability of trained models and enable easy deployment in multi-data-source environments [2], [23]. Industry 5.0 will be fueled by the incredible capabilities provided by 6G that highlight human-centric and sustainable industrial operations. AI and robotics will work together to increase productivity, safety, and flexibility in manufacturing. Through real-time insights powered by AI, rapid response to acute needs will be enabled and waste of resources across industries will also be reduced [11], [21]. Besides, low-latency computing is required for applications like autonomous vehicles, gaming, or robotic control. By combining 6G, AI frameworks, rapid data processing, and decision making will be guaranteed. Speed-optimized LSTM (SP-LSTM) models and RL are likely to have a great impact on avoiding delays to reduce resource consumption [1], [6]. 6G will have intelligent offloading strategies to access the optimal times of executing a computationally intensive task in task management in 6G networks by 6G systems. AI-powered decisions will improve energy efficiency and reduce latency, allowing high performance in both edge and cloud environments [4], [22]. The Tactile Internet, promised by 6G, will also enable remote control of devices at the speed of real-time. This will make the way for applications as remote surgeries, high-precision manufacturing, and virtual games. AI will add to the reliability and responsiveness of Tactile Internet enabled systems, to achieve safe operations [2], [45]. Internet of Everything (IoE) will envision a connected ecosystem in which people, things, and systems talk to each other without any human involvement. This massive network will be managed by AI algorithms at the servant level which will optimize communication links and ensure that no resources are wasted [8], [38]. The concept of IoE goes a step further than IoT by introducing not only devices but also people, processes, and data, working on an integrated network. The following are the main components of IoE:

- People: The interaction of people with devices and systems contributes to providing data and their requests for tailor-made services.
- Devices: Smart devices, sensors, and machines collect and share useful data.
- Data: Information generated by human users as well as devices is processed and analyzed for smart decision-making.
- Processes: Automated workflows and systems ensure seamless interactions between humans, devices, and data.
- Truly autonomous, intelligent, and interoperable, the IoE ecosystem has applications in smart cities, autonomous transport, and personalized health care.

Crucial for the realization of the vision of IoE, the 6G network provides the URLLC, mMTC, and eMBB needed for seamless connectivity. The main conditions of 6G for IoE include ultra-low latency with delays of about 0.1 msec that will enable real-time interaction between devices and systems, combined with massive connectivity supporting billions of devices per square kilometer to ensure comprehensive coverage for all IoE applications Gubbi et al. [122]. Furthermore, these networks will prioritize energy efficiency by targeting power consumption as an optimization factor, allowing environmentally sustainable operation of IoE devices Chowdhury et al. [116]. AI algorithms are critical for managing the IoE networks' complexity, optimizing the communication links, and assuring efficient resource utilization. Key AI-driven capabilities include predictive analytics, where AI analyzes historical and current data to forecast network traffic, device failures, and resource demand, exemplified by its ability to optimize energy consumption in smart grids by anticipating demand patterns. Building on this, autonomous resource allocation enables AI to dynamically distribute bandwidth and power based on real-time conditions, such as prioritizing emergency IoT sensors during network congestion. Furthermore, AI-powered anomaly detection examines IoE data for unusual patterns, facilitating early identification of both security threats like network intrusions and operational issues such as leaks in water distribution systems. 6G networks will also be known for dynamic network reconfiguration since their main target is to immediately cater network conditions. AI will be in charge and observing network states to make online changes to resource allocation, load balancing, or fault recovery. This will enable communication in a resilient and reliable way even in very challenging circumstances [5], [6]. At last, self-learning systems for 6G will perform an autopilot-style continuous improvement, where no human is needed. These systems will be able to use AI to learn from experience, improve performance, and solve new challenges independently. Self-learning capabilities will be commercially beneficial for scenarios in which these models evolve, such as the future of autonomous systems and cities Celik et al. [45].

## VIII. Conclusion

### A. Transformative Role of AI in 6G Networks

The integration of artificial intelligence (AI) into 6G networks represents a paradigm shift from rule-based network control to *AI-native, self-evolving communication systems*. AI-assisted services such as massive machine-type communications (mMTC), ultra-reliable low-latency communications (URLLC), and integrated sensing and communication (ISAC) enable unprecedented levels of automation, intelligence, and adaptability. These capabilities underpin emerging applications including autonomous systems, smart cities, holographic communications, and the tactile Internet, all of which demand ultra-high data rates, extreme reliability, and massive connectivity.

### B. Key Technical Challenges in AI-Enabled 6G

#### 1) Security, Privacy, and Trust

AI-driven network intelligence introduces new attack surfaces, including adversarial learning, data poisoning, and model inversion attacks. Decentralized intelligence further complicates security assurance across heterogeneous devices and administrative domains. While federated learning (FL), blockchain-based privacy mechanisms, and adversarial defense strategies show promise, their scalability, convergence guarantees, and robustness under non-iid data distributions remain open research challenges.

#### 2) Energy Efficiency and Sustainability

The computational complexity of AI model training and inference, combined with ultra-dense device deployments, poses significant sustainability challenges for 6G networks. Without careful design, AI-native architectures could dramatically increase network energy consumption. Consequently, advances in green AI, energy-aware model optimization, energy harvesting, and intelligent power control mechanisms are essential to ensure long-term sustainable operation.

#### 3) Hardware and Infrastructure Constraints

The realization of AI-native 6G requires tight integration between AI-tailored hardware accelerators, edge and cloud orchestration platforms, terahertz (THz) communication systems, and potentially quantum-assisted computing paradigms. However, hardware heterogeneity, deployment cost, and thermal limitations remain critical barriers to large-scale adoption.

### C. Enabling Technologies and Architectural Evolution

#### 1) Programmable and Softwarized Networks

Software-defined networking (SDN) and network function virtualization (NFV) provide the programmability and flexibility required to manage the dynamic and heterogeneous nature of 6G environments. These technologies enable rapid service deployment, adaptive resource allocation, and AI-driven control loops. Nevertheless, achieving interoperability and global standardization across vendors and regions remains a major challenge, requiring coordinated international efforts.

#### 2) Network Intelligence and Analytics

AI-powered network data analytics frameworks enable closed-loop optimization, predictive fault management, and autonomous decision-making. Such capabilities are fundamental to realizing self-optimizing, self-healing, and self-scaling 6G networks capable of adapting to changing traffic demands and environmental conditions in real time.

### D. Ethical and Governance Considerations

Beyond technical performance, the widespread deployment of AI in 6G networks raises important ethical and governance concerns. Ensuring fairness, accountability, transparency, and non-discriminatory access to network services is essential for societal acceptance. Explainable AI (XAI) and quality-of-trust and ethics (QoTE) frameworks play a critical role in enhancing transparency, enabling auditability, and aligning AI-driven decisions with societal values and regulatory requirements.

### E. Autonomous and Privacy-Preserving 6G Operation

Self-learning and self-reconfiguring mechanisms will enable autonomous resource management, resilience, and efficiency under highly dynamic network conditions. User privacy preservation remains a fundamental requirement. Rather than centralized raw data aggregation, federated learning naturally aligns with the distributed architecture of 6G networks, enabling collaborative intelligence while keeping sensitive data local to devices.

### F. Distinct Contributions of This Survey

This survey distinguishes itself through a practical, cross-layer perspective. Rather than focusing solely on isolated AI algorithms or specific application domains, it systematically connects service requirements with AI-driven network analytics and real-world deployment constraints, including hardware limitations, energy consumption, and system interoperability. Furthermore, governance and ethical considerations are examined alongside technical enablers such as federated learning, reinforcement learning, and advanced communication technologies. This holistic approach provides actionable insights for prototype development and early-stage 6G deployments.

**TABLE 10.** Key Challenges and Open Research Issues in AI-Enabled 6G

| Category | Challenge | Open Research Issues |
| --- | --- | --- |
| Security | Adversarial AI | Robustness, scalability |
| Energy | AI computation cost | Green AI, power-aware models |
| Hardware | THz and edge devices | Cost, thermal constraints |
| Ethics | Trust and fairness | XAI, QoTE frameworks |

**TABLE 11.** Key Enabling Technologies for AI-Native 6G

| Technology | Role in 6G | Key Benefits |
| --- | --- | --- |
| Federated Learning | Distributed intelligence | Privacy preservation |
| SDN/NFV | Network programmability | Flexibility, scalability |
| THz Communication | Extreme data rates | New spectrum access |
| Explainable AI | Transparency | Trust and accountability |

### G. Final Remarks

The convergence of AI and 6G networks will fundamentally redefine wireless connectivity and digital interaction. Realizing this vision requires addressing intertwined challenges spanning security, sustainability, hardware innovation, standardization, and ethics. Continued collaboration among academia, industry, and regulatory bodies will be essential in shaping trustworthy, efficient, and intelligent AI-powered 6G networks. The insights presented in this survey aim to serve as a roadmap for future research and development toward this goal.

## IX. Future Research Directions in AI for 6G

Despite the significant progress already achieved in the integration of AI into 6G networks, several open challenges and unexplored opportunities remain. Addressing these issues will be essential to unlock the full potential of intelligent 6G systems and to ensure their applicability in real-world scenarios.

First, future work should focus on AI-driven self-optimization of 6G networks, where adaptive algorithms can dynamically manage spectrum, energy, and latency constraints in real time. RL and other adaptive techniques have strong potential but require further exploration to meet stringent quality-of-service (QoS) demands.

Second, the role of explainable AI (XAI) in 6G must be emphasized. As 6G is expected to support mission-critical applications such as autonomous driving and remote surgery, transparent decision-making becomes crucial. Future research should investigate lightweight and interpretable AI models that can balance accuracy, explainability, and computational efficiency.

Third, edge intelligence and FL represent promising avenues for distributing AI processing closer to users. By enabling privacy-preserving and energy-efficient learning at the network edge, these approaches could reduce dependency on centralized cloud infrastructures. However, scalable algorithms, robust communication protocols, and efficient resource allocation strategies remain largely open research questions.

Fourth, AI-enabled security and privacy will be a cornerstone of 6G reliability. As AI itself becomes a target for adversarial attacks, research into robust and trustworthy AI mechanisms will be essential. Integrating AI with technologies such as blockchain may offer decentralized and secure solutions for data protection and trust management.

Finally, future investigations should explore the synergy between AI and emerging paradigms, including quantum computing for accelerated optimization and holographic or tactile internet services for immersive communication. Such cross-disciplinary research could define the next frontier of 6G capabilities.

In summary, advancing AI research for 6G requires a multi-dimensional approach, spanning from algorithmic innovation to system-level integration. By addressing these open questions, the research community can provide a roadmap for the realization of scalable, secure, and intelligent 6G networks.

## X. List of Abbreviations

1) **3D** – Three-Dimensional
2) **5G** – Fifth Generation
3) **6G** – Sixth Generation
4) **AD** – Anomaly Detection
5) **AI** – Artificial Intelligence
6) **AI-RAN** – AI-driven Radio Access Network
7) **AMP** – Approximate Message Passing
8) **API** – Application Programming Interface
9) **AR** – Augmented Reality
10) **ASIC** – Application-Specific Integrated Circuit
11) **B5G** – Beyond Fifth Generation
12) **BFT** – Byzantine Fault Tolerance
13) **BS** – Base Station
14) **CIDS** – Collaborative Intrusion Detection System
15) **CNN** – Convolutional Neural Network
16) **CSI** – Channel State Information
17) **CTDE** – Centralized Training with Decentralized Execution
18) **DDPG** – Deep Deterministic Policy Gradient
19) **DL** – Deep Learning
20) **DLT** – Distributed Ledger Technology
21) **DQN** – Deep Q-Network
22) **DRL** – Deep Reinforcement Learning
23) **DT** – Digital Twin
24) **eMBB** – Enhanced Mobile Broadband
25) **EMF** – Electromagnetic Field
26) **FedAvg** – Federated Averaging
27) **FedProx** – Federated Proximal Algorithm
28) **FEL** – Federated Edge Learning
29) **FL** – Federated Learning
30) **FPGA** – Field-Programmable Gate Array
31) **GAN** – Generative Adversarial Network
32) **GDPR** – General Data Protection Regulation
33) **GenAI** – Generative Artificial Intelligence
34) **GNN** – Graph Neural Network
35) **GPU** – Graphics Processing Unit
36) **gNB** – Next-Generation Node Base (5G/6G Base Station)
37) **HARQ** – Hybrid Automatic Repeat Request
38) **ICNIRP** – International Commission on Non-Ionizing Radiation Protection
39) **IDS** – Intrusion Detection System
40) **IIoT** – Industrial Internet of Things
41) **IoE** – Internet of Everything
42) **IoT** – Internet of Things
43) **ISAC** – Integrated Sensing and Communication
44) **ISTA** – Iterative Shrinkage-Thresholding Algorithm
45) **KPI** – Key Performance Indicator
46) **LIS** – Large Intelligent Surface
47) **LLM** – Large Language Model
48) **LPA** – Learning Policy Agent
49) **LSTM** – Long Short-Term Memory
50) **MARL** – Multi-Agent Reinforcement Learning
51) **MIMO** – Multiple-Input Multiple-Output
52) **ML** – Machine Learning
53) **mMTC** – Massive Machine-Type Communication
54) **NFV** – Network Function Virtualization
55) **NOMA** – Non-Orthogonal Multiple Access
56) **NTN** – Non-Terrestrial Network
57) **NWDAF** – Network Data Analytics Function
58) **O-RAN** – Open Radio Access Network
59) **OFDM** – Orthogonal Frequency-Division Multiplexing
60) **PHY** – Physical Layer
61) **PoA** – Proof of Authority
62) **POMDP** – Partially Observable Markov Decision Process
63) **PoS** – Proof of Stake
64) **PoW** – Proof of Work
65) **PPO** – Proximal Policy Optimization
66) **PRB** – Physical Resource Block
67) **QAOA** – Quantum Approximate Optimization Algorithm
68) **QKD** – Quantum Key Distribution
69) **QoE** – Quality of Experience
70) **QoS** – Quality of Service
71) **QoTE** – Quality of Trust Experience
72) **RAN** – Radio Access Network
73) **RF** – Radio Frequency
74) **RIC** – RAN Intelligent Controller
75) **RL** – Reinforcement Learning
76) **RNN** – Recurrent Neural Network
77) **RIS** – Reconfigurable Intelligent Surface
78) **RRM** – Radio Resource Management
79) **RT** – Real Time
80) **SDA** – Software-Defined Architecture
81) **SDN** – Software-Defined Networking
82) **SLA** – Service Level Agreement
83) **SMPC** – Secure Multi-Party Computation
84) **SP-LSTM** – Speed-Optimized Long Short-Term Memory
85) **SSL** – Self-Supervised Learning
86) **TD3** – Twin Delayed Deep Deterministic Policy Gradient
87) **THz** – Terahertz
88) **TI** – Tactile Internet
89) **TPU** – Tensor Processing Unit
90) **TTI** – Transmission Time Interval
91) **UE** – User Equipment
92) **UHD** – Ultra-High Definition
93) **URLLC** – Ultra-Reliable Low-Latency Communication
94) **V2X** – Vehicle-to-Everything Communication
95) **VAE** – Variational Autoencoder
96) **VR** – Virtual Reality
97) **WPT** – Wireless Power Transfer
98) **XAI** – Explainable Artificial Intelligence
99) **XR** – Extended Reality

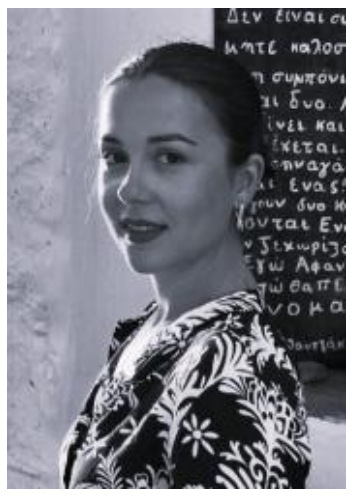

**CONSTANTINA CHATZIELEFTHERIOU** received the B.Sc. degree in Informatics and Telecommunications and the M.Sc. degree in Computer, Telecommunications, and Network Engineering from the National and Kapodistrian University of Athens, Greece, in 2017 and 2020, respectively. She also completed studies in Pedagogy and Teaching Competence at the Hellenic Open University, Patras, Greece, in 2023. She has served as a Teaching Assistant at the National and Kapodistrian University of Athens and currently works as a private tutor and freelancer in Athens, Greece. Her research interests include software engineering, 6G networks, wireless communication, artificial intelligence, machine learning, the Internet of Things (IoT), cybersecurity, and sustainable development. She is currently pursuing her Ph.D. at the Harokopio University of Athens, under the title "Development and Optimization of 6G Networks using Artificial Intelligence."

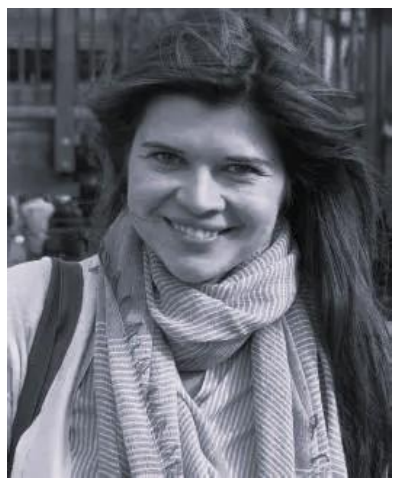

**EIRINI LIOTOU** is an Assistant Professor at the Department of Informatics and Telematics of Harokopio University of Athens since 2024. She holds a PhD degree from the Department of Informatics and Telecommunications at the National and Kapodistrian University of Athens. She obtained the MSc in Informatics and Telecommunications from the same university, the MSc in Communications and Signal Processing from the Imperial College of London, and the Diploma in Electrical and Computer Engineering from the National Technical University of Athens. She has worked as a Senior Software Engineer at Siemens Enterprise Communications in the RD department. She has also worked as a Post-Doc researcher in the Department of Informatics and Telecommunications at the National and Kapodistrian University of Athens, and as a Scientific Project Manager for EU Research Projects at the Institute of Communications and Computer Systems (ICCS). She has authored more than 30 peer-reviewed publications, and she has participated in numerous research and development programs. She also serves as a Reviewer for EU projects (as an EC expert). Her main research interests revolve around 6G mobile networks, with a particular focus on Software-Defined Networking (SDN), Network Functions Virtualization (NFV), AI in networking and network emulation, as well as Software-Defined Vehicle (SDV) and V2X communications.